%% file: massless.tex
\documentclass[paper=A4,pagesize]{scrartcl}

\input{preamble}

\title{On the analytic computation of massless propagators in dimensional regularization}
\author{Erik Panzer\footnote{\href{mailto:panzer@mathematik.hu-berlin.de}{\nolinkurl{panzer@mathematik.hu-berlin.de}}}}
\date{\today}

\hypersetup{%
	pdftitle = {On the analytic computation of massless propagators in dimensional regularization},
	pdfauthor = {Erik Panzer}
}

\begin{document}

\maketitle

\begin{abstract}
	We comment on the algorithm to compute periods using hyperlogarithms, applied to massless Feynman integrals in the parametric representation. Explicitly, we give results for all three-loop propagators with arbitrary insertions including order $\varepsilon^4$ and show examples at four and more loops.

	Further we prove that all coefficients of the $\varepsilon$-expansion of these integrals are rational linear combinations of multiple zeta values and in some cases possibly also alternating Euler sums.
\end{abstract}

\section{Introduction and results}
We consider Feynman graphs $G$ of propagator type (having only two external legs carrying a momentum $q$) and their associated dimensionally regularized \cite{Collins} massless scalar\footnote{We comment on tensor integrals in section~\ref{sec:tensors}.} Euclidean Feynman integral
\begin{equation}
	\Phi_{G}\left(\ep_1,\ldots,\ep_E,\D; q^2\right)
	\defas
	\prod_{i=1}^h \int \frac{\dd[\D] k_i}{\pi^{\D/2}}
	\prod_{e \in E} \frac{1}{p_e^{2\ep_e}},
	\label{eq:FI-impulsraum}
\end{equation}
were $E$ denotes the edges and $h$ the number of loops in $G$. Here we fixed the dimension $\D=4 - 2\varepsilon$ and allow for arbitrary powers $\ep_e = 1 + \varepsilon \epe_e$ of the propagators $p_e^2$. Recall that $p_e$ is a linear combination of $q$ and the loop momenta $k_i$ as dictated by the choice of a basis of loops and momentum conservation at each vertex.

\begin{figure}
	\begin{gather*}
		L = \Graph{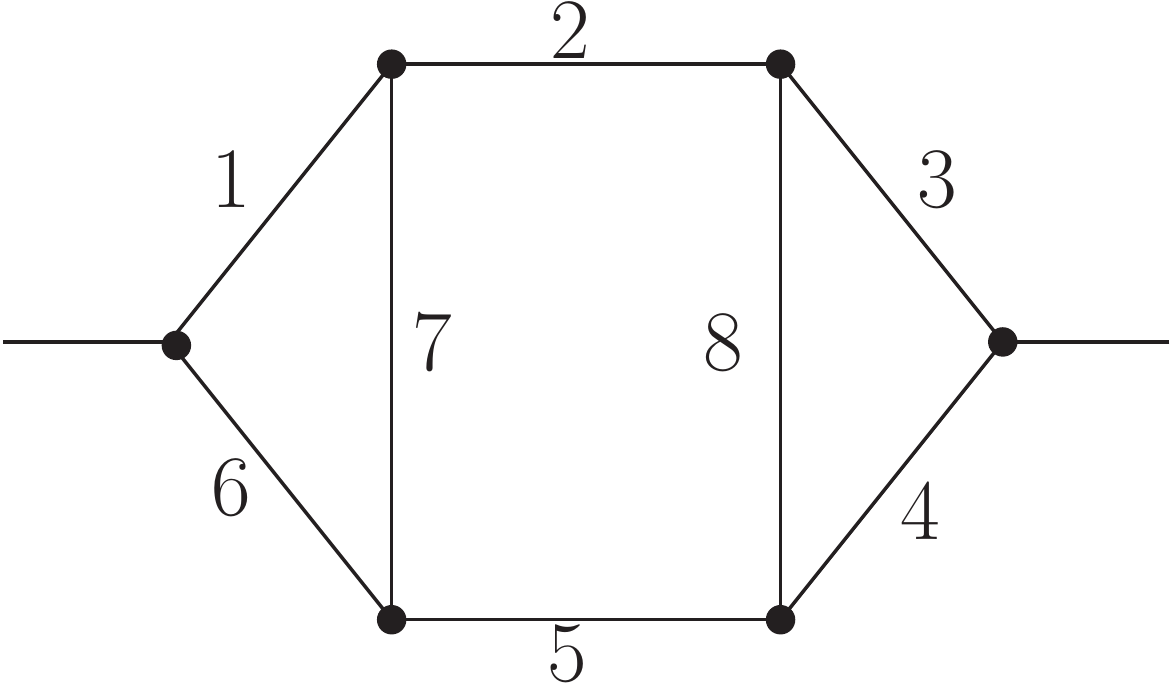}
		\qquad
		N = \Graph{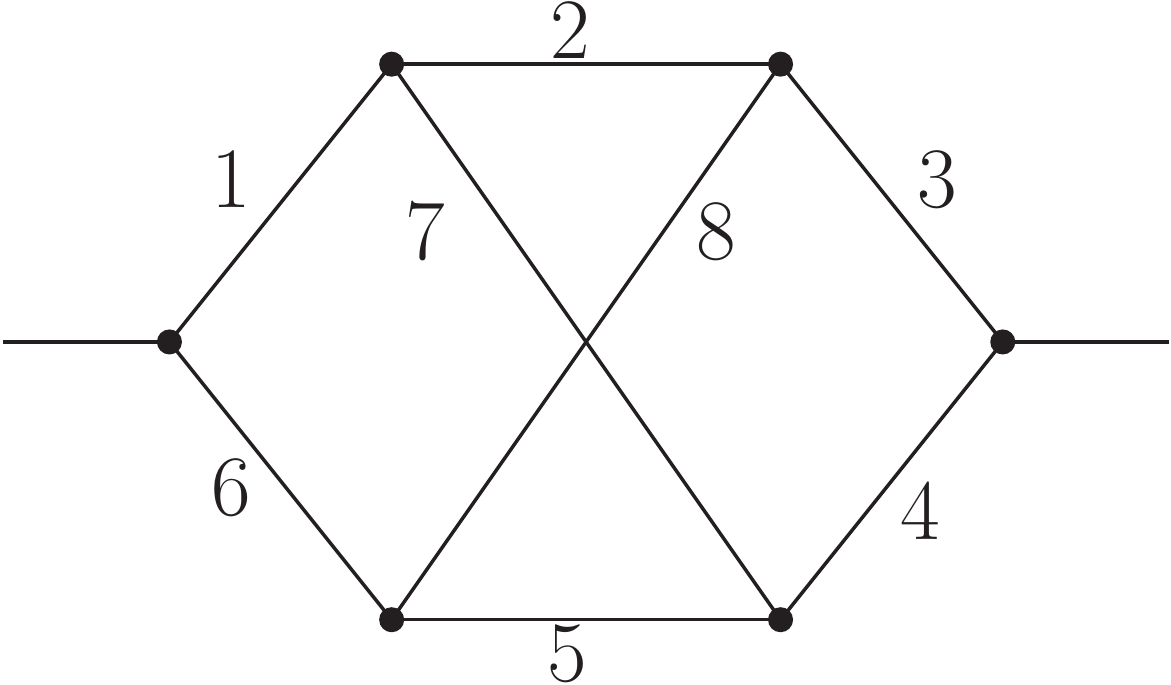} \\
		M = \Graph{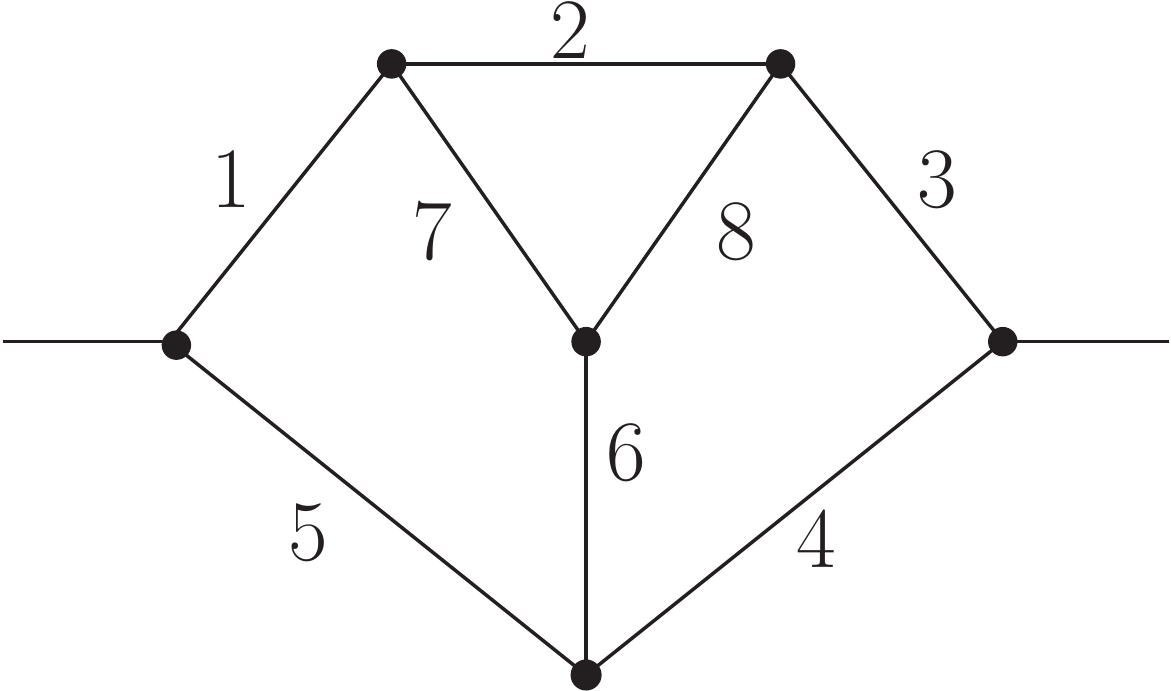}
		\qquad
		V = \Graph{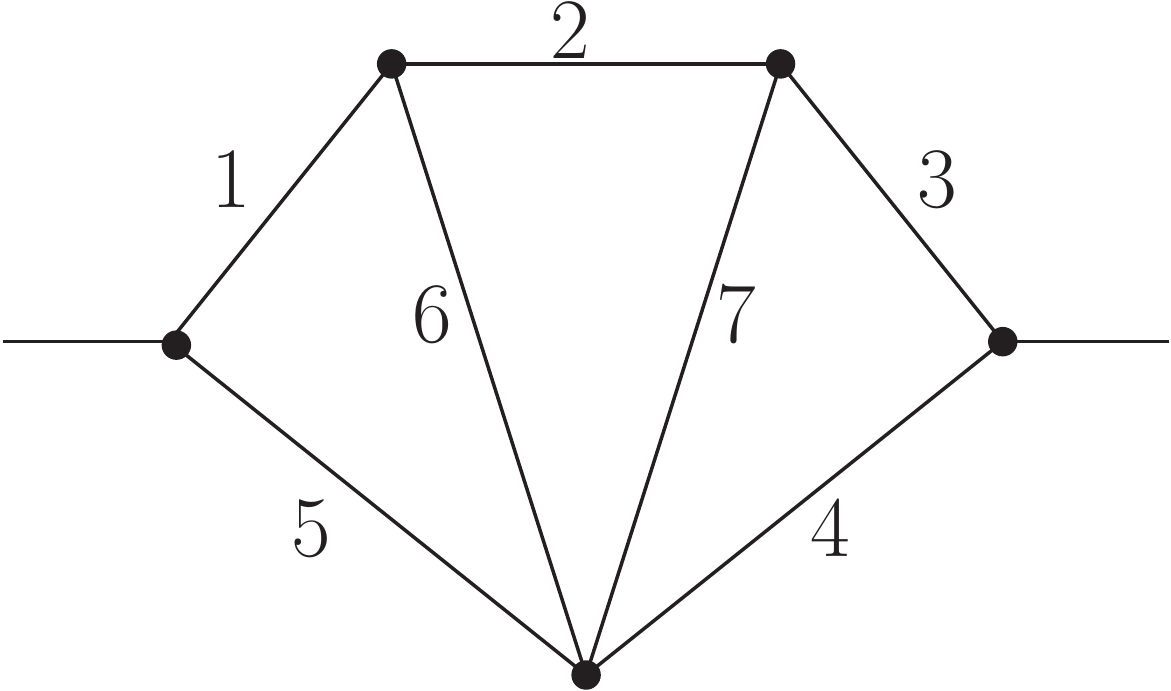}
		\qquad
		Q = \Graph{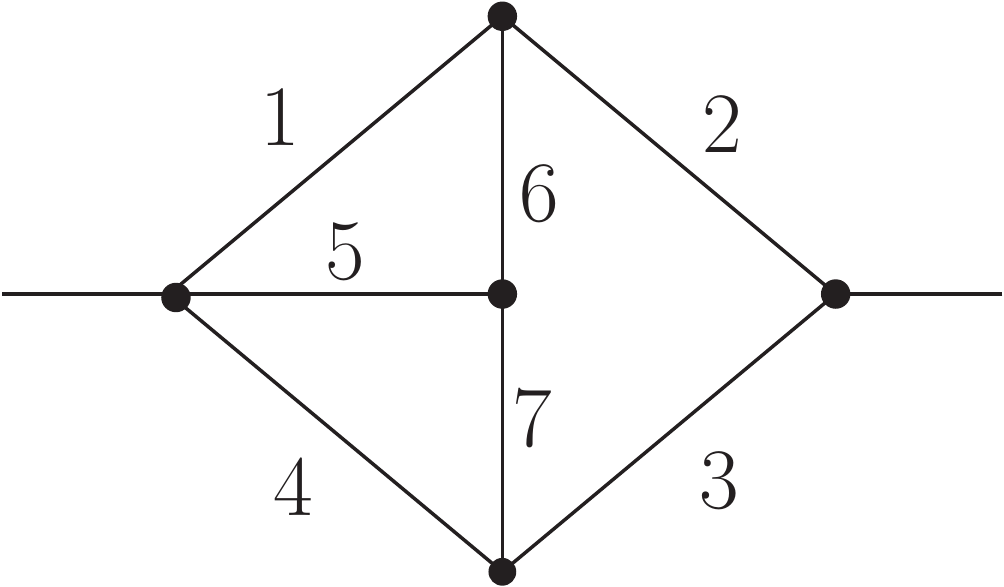}
	\end{gather*}
	\caption{The five different three-loop $p$-integrals were classified in \cite{ChetyrkinTkachov:IBP}.}
	\label{fig:3loops}
\end{figure}

Sometimes referred to as \emph{$p$-integrals} \cite{BaikovChetyrkin:FourLoopPropagatorsAlgebraic}, these currently form a major tool for perturbative calculations in {\qft} and much effort is being invested to compute individual terms of their \emph{$\varepsilon$-expansion} $\Phi_{G} \in \R[\varepsilon^{-1}, \varepsilon]]$ defined as the Laurent series of \eqref{eq:FI-impulsraum} at $\D\rightarrow 4$. 
However, only for the case $h=1$ of a single loop where
\begin{equation}
	G(\ep_1,\ep_2)
	\defas
	\Phi_{\Graph{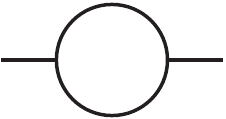}}(\ep_1,\ep_2,\D;q^2)
	= q^{-2\varepsilon} \frac{\Gamma\left( \frac{\D}{2} - \ep_1 \right) \Gamma\left( \frac{\D}{2} - \ep_2 \right) \Gamma\left( \ep_1+\ep_2 - \frac{\D}{2} \right)}{\Gamma(\ep_1)\Gamma(\ep_2)\Gamma(\D-\ep_1-\ep_2)}
	\label{eq:oneloop}
\end{equation}
and $h=2$ \cite{BierenbaumWeinzierl:TwoPoint} these expansions are analytically known or computable. 
Already in the three-loop case $h=3$ available results restrict to low orders in $\varepsilon$ and special assumptions of the form $\epe_e=0$ (that is $\ep_e=1$) for some edges $e$ or \emph{uniqueness} relations \cite{Kazakov:Uniqueness} need to be imposed. 
For a striking example note that only the first three coefficients\footnote{We present results with the prefactor $G_0^h$ where $G_0 \defas \varepsilon G(1,1)$ to ease comparison with the \emph{G-scheme} employed in other publications. It also completely absorbs any Euler-Mascheroni constants $\gamma$.} of
\begin{gather}
	\frac{\Phi_N(1,\ldots,1,4-2\varepsilon; 1)}{G_0^3 (1-2\varepsilon)^2}
	= %
\input{Expansions/N-00000000}
	\label{eq:N-expansion}
\end{gather}
have so far been known analytically (high-precision numeric approximations are available in \cite{Bekavac:MasslessIntegralsHarmonicSums,LeeSmirnov:EasyWay}), where $N$ denotes the non-planar propagator of figure~\ref{fig:3loops}.
The $\varepsilon^2$-contribution was only determined recently \cite{BaikovChetyrkin:FourLoopPropagatorsAlgebraic} in a very indirect way, simultaneously considering many different $p$-integrals and relations between their coefficients that arise from integration-by-parts identities (\IBP) and the \emph{Glue-and-Cut} (\GaC) symmetry.
It therefore seems that the presently employed techniques for analytic evaluation of $p$-integrals are rather limited.

\begin{figure}
	\begin{equation*}
		Y_3 = \CloseProp{L} = \CloseProp{M} = \Graph{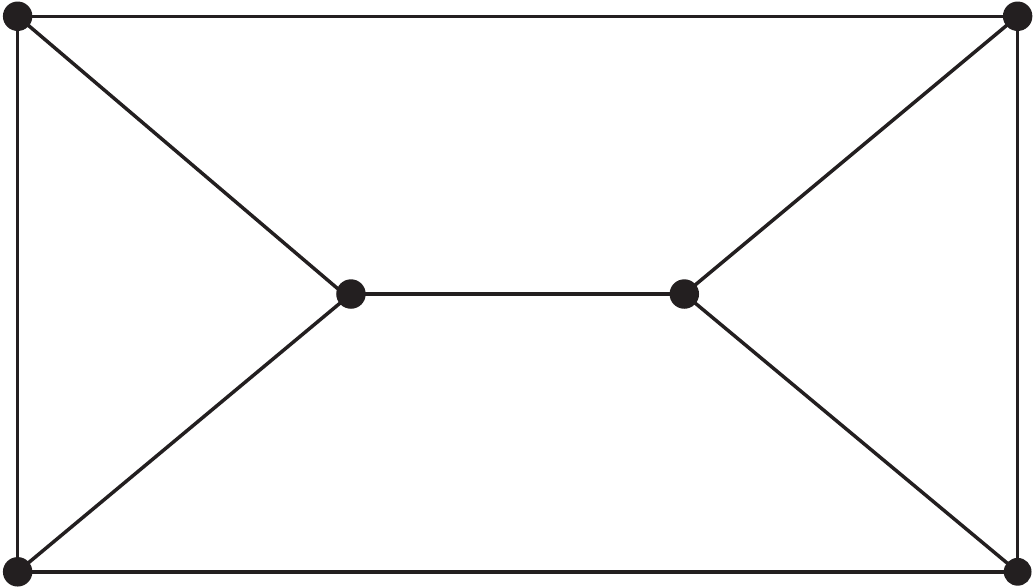}
		\qquad
		W_4 = \CloseProp{Q} = \CloseProp{V} = \Graph{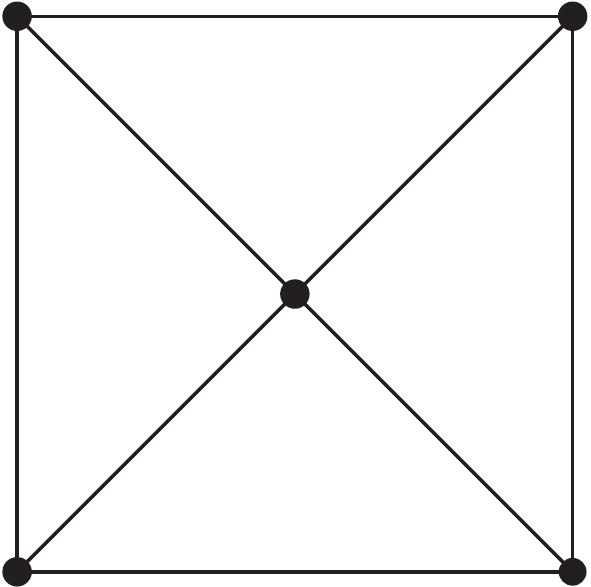}
		\qquad
		K_{3,3} = \CloseProp{N} = \Graph{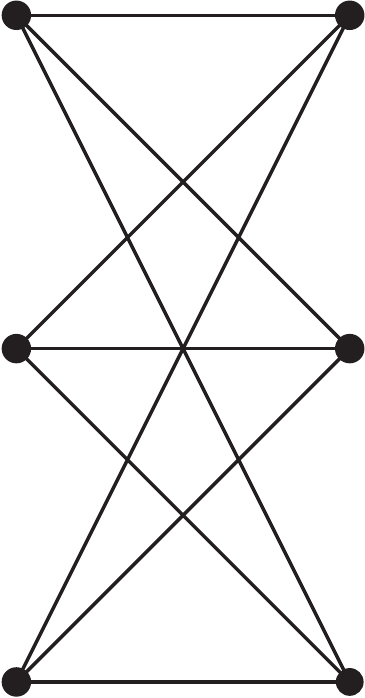}
	\end{equation*}
	\caption{Glueing the external edges of $L$ and $M$ gives the triangular prism $Y_3$, while $Q$ and $V$ yield the wheel with four spokes $W_4$ and $N$ results in the complete bipartite graph $K_{3,3}$.}
	\label{fig:3loops-completions}
\end{figure}

However, in \cite{Brown:TwoPoint} Francis Brown developed an algorithm very well suited to compute this kind of integrals.
It can be applied to a graph $G$ if it is \emph{linearly reducible} (see definition~\ref{def:linearly-reducible}) and restricts the periods that may occur in the result. Our purpose is twofold:

\paragraph{All-order constraints on periods:}
The analysis of \cite{Brown:TwoPoint} proved that all coefficients of the $\varepsilon$-expansion are (rational linear combinations of) multiple zeta values (\MZV)
\begin{equation}
	\zeta_{n_1,\ldots,n_d}
	= \sum_{1 \leq k_1 < \ldots < k_d} \frac{1}{k_1^{n_1}\ldots k_d^{n_d}},
	\qquad
	n_1,\ldots,n_d \in \N
	\quad\text{with}\quad
	n_d \geq 2,
	\label{eq:MZV}
\end{equation}
in the case of the three-loop propagator graphs $Q$ and $V$ of figure~\ref{fig:3loops}.
We extend this consideration to all three-loop graphs in
\begin{theorem}%
	\label{theorem:3loops}%
	All three-loop graphs of figure~\ref{fig:3loops} are linearly reducible. 
	Every coefficient of their $\varepsilon$-expansions $G_0^{-3} \Phi$ is a rational linear combination of {\MZV} for the planar graphs $L$, $M$, $Q$ and $V$.  For the non-planar graph $N$ also alternating Euler sums may appear.
\end{theorem}
It was shown in \cite{Brown:TwoPoint} that propagators $G$ can be calculated from periods of the vacuum graph $\CloseProp{G}$ obtained by glueing the external legs to merge into a new internal edge, as depicted in figure~\ref{fig:3loops-completions}.
Therefore it becomes most efficient to study these instead of the propagators themselves.

\begin{figure}
	\begin{gather*}
		{_5 P_1} = \Graph{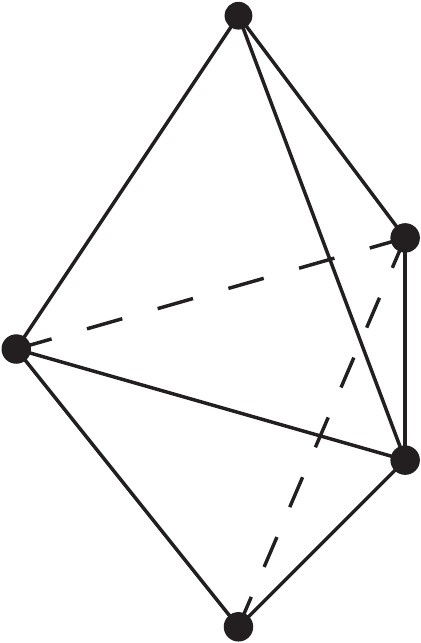}
		\quad
		{_5 P_2} = \Graph{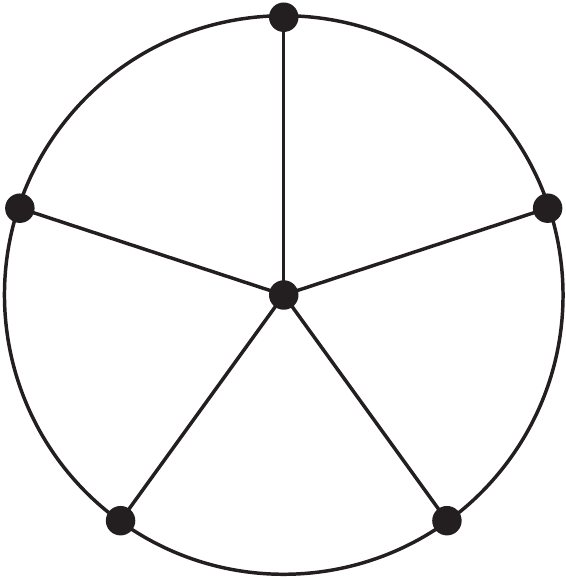}
		\quad
		{_5 P_3} = \Graph{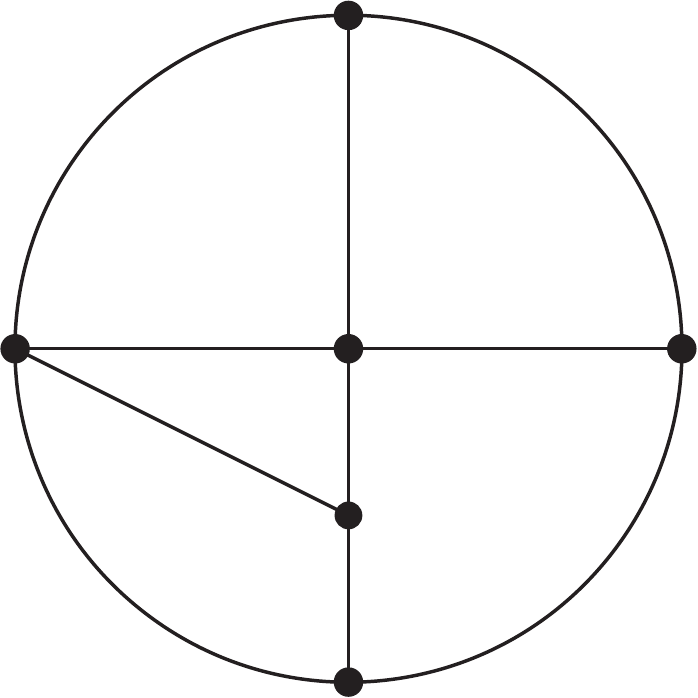}
		\quad
		{_5 P_4} = \Graph{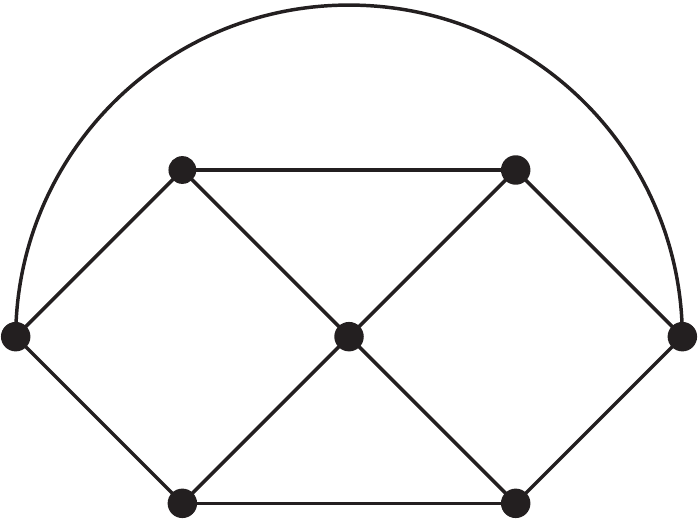} \\
		{_5 P_5} = \Graph{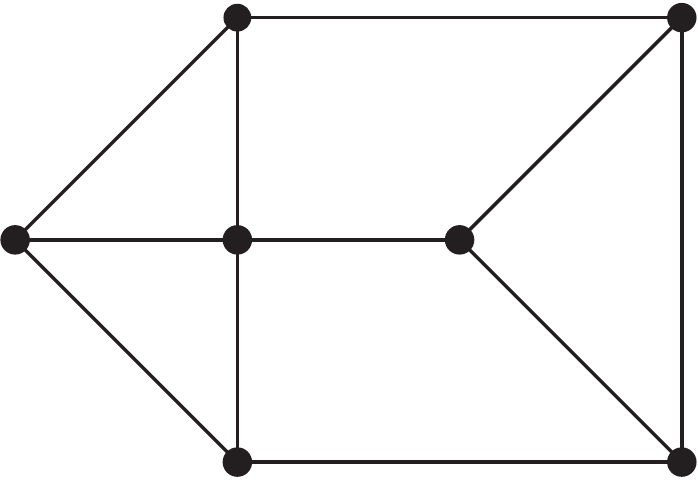}
		\quad
		{_5 P_6} = \Graph{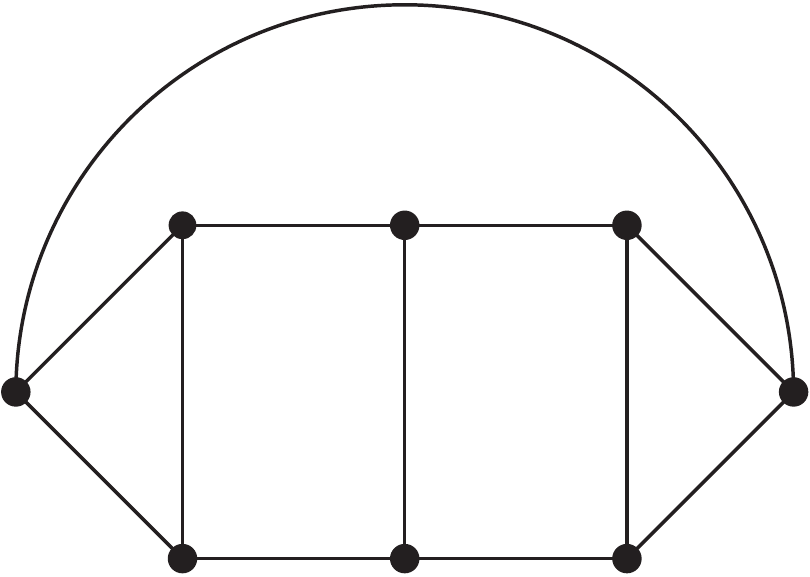}
		\quad
		{_5 P_7} = \Graph{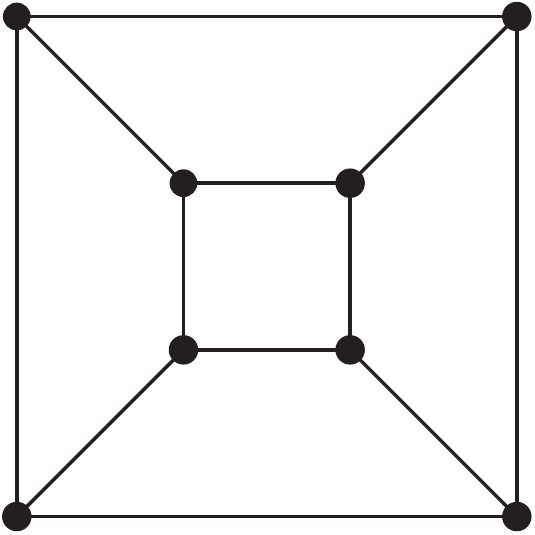} \\
		{_5 N_1} = \Graph{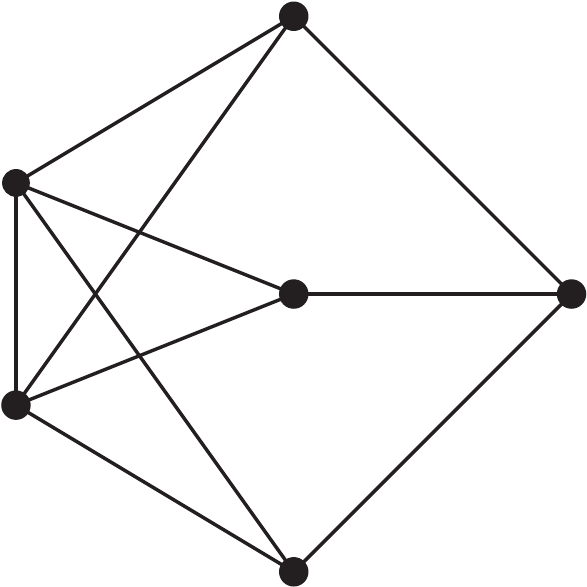}
		\quad
		{_5 N_2} = \Graph{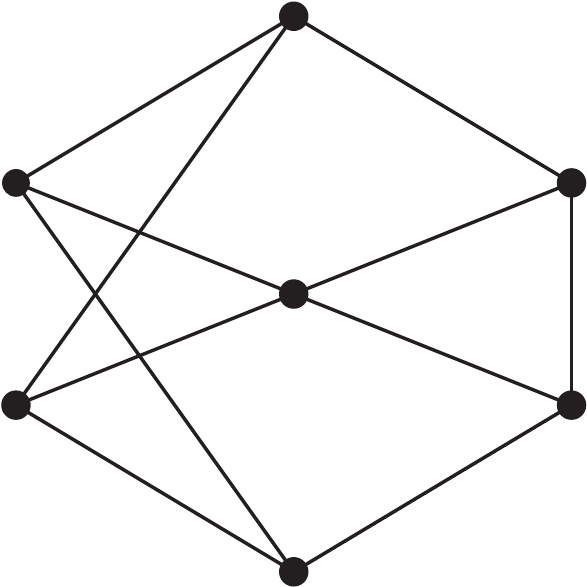}
		\quad
		{_5 N_3} = \Graph{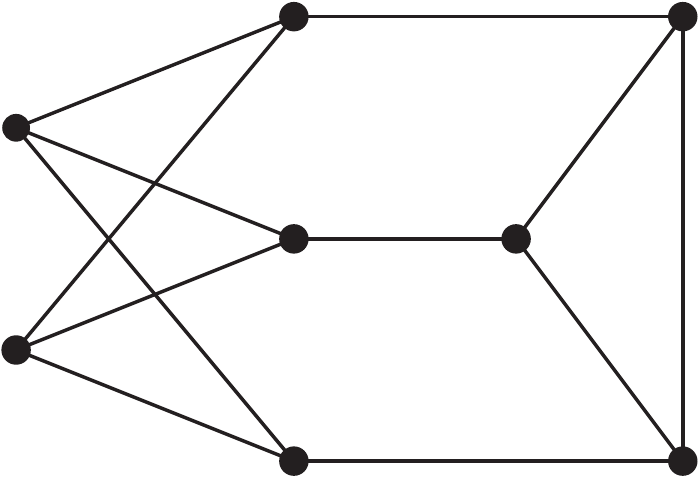}
		\quad
		{_5 N_4} = \Graph{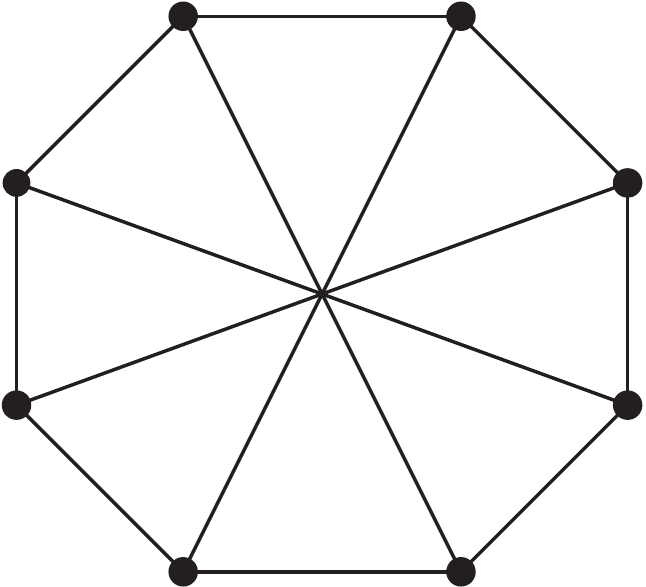}
	\end{gather*}
	\caption{%
	All five-loop graphs without one-scale subgraphs, divided into planar ($P$) and non-planar ($N$) ones. 
	The Zig-zag graph ${_5 P_3}$ and ${_5 N_1}$ ($K_{3,3}$ with an additional edge) were considered in \cite{Brown:TwoPoint}. 
	Cutting any edge produces a propagator graph with four loops. 
	The master integrals of \cite{BaikovChetyrkin:FourLoopPropagatorsAlgebraic,LeeSmirnov:FourLoopPropagatorsWeightTwelve}, some shown in figure~\ref{fig:4loops}, give $\CloseProp{M_{3,5}} = \CloseProp{M_{3,6}} = {_5 P_1}$ (the complete graph $K_5$ minus one edge), $\CloseProp{M_{4,4}} = {_5 P_3}$, $\CloseProp{M_{4,5}} = {_5 N_1}$, $\CloseProp{M_{5,1}} = {_5 N_2}$, $\CloseProp{M_{6,1}} = {_5 P_7}$ (the cube) and $\CloseProp{M_{6,2}} = \CloseProp{M_{6,3}} = {_5 N_4}$.}
	\label{fig:4loops-completions}
\end{figure}

We can now state our analysis of the four-loop case in
\begin{theorem}%
	\label{theorem:4loops}%
	All four-loop propagators are linearly reducible and arise upon cutting one edge of one of the five-loop graphs of figure~\ref{fig:4loops-completions}. 
	The $\varepsilon$-expansions $G_0^{-4}\Phi$ of such propagators without subdivergences are rational linear combinations of {\MZV}, except possibly for those that are cuts of the non-planar graphs ${_5 N_1}, {_5 N_2}, {_5 N_3}, {_5 N_4}$ or ${_5 P_7}$, in which case alternating Euler sums may appear.
\end{theorem}
This was already proved for ${_5 P_3}$ in \cite{Brown:TwoPoint} which also considered ${_5 N_1}$ but could not reveal its linear reducibility. 
As a consequence the three distinct non-planar propagators arising by cutting one edge of this graph could have entailed more complicated periods (namely multiple polylogarithms at sixth roots of unity), which had been looked for numerically in \cite{LeeSmirnov:FourLoopNonPlanarPropagators} to no avail. This suggested the sufficiency of {\MZV} and motivated the quest for lower bounds on the periods that resulted in theorem~\ref{theorem:4loops} above.

We remark that propagators with subdivergences can be treated with our method as well, but they require a separate inspection as commented on in section~\ref{sec:subdivergences}.

\paragraph{Practical considerations and explicit calculations:}
The above results imply that we can compute these $p$-integrals analytically using hyperlogarithms, in principle to arbitrary order in $\varepsilon$ (though in practice it will be bounded by memory and time constraints as well as the efficiency of the implementation).
We programmed this routine in the computer algebra system {\Maple}, closely following the description in \cite{Brown:TwoPoint}.

After introducing the parametric representation and polynomial reduction in section~\ref{sec:parametric}, we report our complete three- and some four-loop results in sections~\ref{sec:3loops} and \ref{sec:4loops}. Some examples at five and six loops follow in section~\ref{sec:moreloops} before we close with some remarks in section~\ref{sec:comments}.

To our knowledge, most of the periods we calculated have so far been either unknown or were only conjectured based on examination of high precision numerical approximations as reported on in \cite{SmirnovTentyukov:FourLoopPropagatorsNumeric,LeeSmirnov:FourLoopNonPlanarPropagators,LeeSmirnov:FourLoopPropagatorsWeightTwelve}.

For completeness and further use all of our results may be obtained in computer-readable form from \cite{Panzer:MasslessPropagatorsData} and are also attached as ancillary files to this arXiv submission.

\subsection{Acknowledgements}
The author is indebted to Francis Brown (supported through ERC grant 257638) for hospitality at IHES, his most beautifully written papers, illuminating discussions and his patience to answer many questions of mine.
Oliver Schnetz provided great stimulus and the program \cite{Schnetz:ZetaProcedures} to reduce {\MZV} to a basis.
I like to thank Dirk Kreimer whose Alexander von Humboldt-working group I so much enjoy to be part of for encouragement and not getting tired of asking me to write up these results.

Christian Bogner kindly answered many questions on iterated integrals and checked some periods by an independent program, while Martin L\"{u}ders provided tests of the polynomial reduction through a separate implementation of the compatibility graph method.

Finally I am grateful to my family and friends for moral support and lovely moments free of any thoughts about Feynman diagrams\ldots

\section{Parametric integration}
\label{sec:parametric}

Our approach works in the \emph{parametric representation} \cite{ItzyksonZuber} instead of \eqref{eq:FI-impulsraum}, namely
\begin{align}
	\Phi_{G}\left( \ep_1,\ldots,\ep_E,\D;q^2 \right)
	=& \prod_{e\in E} \int\limits_0^{\infty} \frac{\dd \SP_e\ \SP_e^{\ep_e-1}}{\Gamma(\ep_e)}
	\frac{\exp\left[ -q^2 \frac{\phipol}{\psipol}\right]}{\psipol^{\D/2}}	\label{eq:FI-parametric-affine}
\hide{%
	&= q^{-2\sdc}
		\frac{\Gamma(\sdc)}{\prod_{e\in E}\Gamma(\ep_e)}
		\int \frac{\Omega}{\psipol^{D/2}}
		\left( \frac{\psipol}{\phipol} \right)^{\sdc}
		\prod_{e\in E} \left( 
			\frac{\SP_e \psipol}{\phipol}
		\right)^{\ep_e-1}\\
}%
	= q^{-2\sdc}
		\frac{\Gamma(\sdc)}{\prod_{e\in E}\Gamma(\ep_e)}
		I_{G} \\
	\text{for}\quad I_{G}
	\defas&
		\int \frac{\Omega}{\psipol^{2}}
		\left( \frac{\psipol}{\phipol} \right)^{\abs{E} - 2h}
		\left( \frac{\psipol^{h+1}}{\phipol^h} \right)^{\varepsilon}
		\prod_{e\in E} \left( 
			\frac{\SP_e \psipol}{\phipol}
		\right)^{\varepsilon \epe_e}
	\label{eq:FI-parametric-projective}
\end{align}
with \emph{Schwinger parameters} $\SP_e$ and the negative power counting degree of divergence
\begin{equation}%
	\label{eq:sdc}%
	\sdc
	\defas \sum_{e\in E}\ep_e - \frac{\D}{2} h
	= \abs{E} - 2 h + \varepsilon \left( h + \sum_{e \in E} \epe_e \right).
\end{equation}
While $\Omega$ denotes the projective volume form, we will always evaluate \eqref{eq:FI-parametric-projective} affinely by setting $\SP_e = 1$ for a distinguished edge $e$, leaving $\abs{E}-1$ integrations to be done.
The homogeneous Symanzik polynomials $\psipol,\phipol \in \Z[\setexp{\SP_e}{e \in E}]$ of degrees $h$ and $h+1$ are given by sums
\begin{equation}%
	\label{eq:symanzik}%
	\psipol = \sum_T \prod_{e \notin T} \SP_e
	\quad\text{and}\quad
	\phipol = \sum_F \prod_{e \notin F} \SP_e
\end{equation}
over the spanning trees $T$ or spanning $2$-forests $F$ (separating the two external legs).
\begin{example}%
	\label{ex:Phi_N-parametric}%
	The graph $N$ of figure~\ref{fig:3loops} is convergent with $\sdc = 2 + \varepsilon ( 3 + \sum \epe_e)$, so
	\begin{equation}
		\Phi_N = 
		\left( q^2 \right)^{-\sdc}
		\frac{\Gamma(2 + \varepsilon[3 + \sum \epe_e])}{\prod_{e\in E}\Gamma(1 + \varepsilon \epe_e)}
		\int \frac{\Omega}{\phipol^{2}}
		\left( \frac{\psipol^4}{\phipol^3} \right)^{\varepsilon}
		\prod_{e\in E} \left( 
			\frac{\SP_e \psipol}{\phipol}
		\right)^{\varepsilon\epe_e}.
		\label{eq:Phi_N-parametric}
	\end{equation}
\end{example}
From now on we set $q^2 = 1$ and focus on $I_{G}$, as the explicit $\Gamma$-functions appearing in \eqref{eq:FI-parametric-affine} are immediately expanded using $z\Gamma(z) = \Gamma(z+1)$ and
\begin{equation}
	\Gamma(1-z) = \exp \left[ z \gamma + \sum_{n \geq 2} \frac{\zeta_n}{n} z^n \right]
	\quad\text{for}\quad
	\abs{z} < 1.
	\label{eq:Gamma-expansion}
\end{equation}
Apart from a rational prefactor, these contribute only {\MZV} to $\Phi_G$ and a factor $e^{-\varepsilon\gamma h}$ that is absorbed by pulling out the prefactor $G_0^h = \left[\varepsilon G(1,1)\right]^h$.

\subsection{Polynomial reduction}%
\label{sec:reduction}
If $G$ is \emph{primitive} (free of subdivergences), the projective integral in \eqref{eq:FI-parametric-projective} converges and we can compute its $\varepsilon$-expansion in the form
\begin{equation}
	I_{G}
		= \sum_{k,k_1,\ldots,k_E}
			\frac{\varepsilon^{k+k_1+\ldots+k_E}}{k! k_1! \ldots k_E!} \prod_{e\in E} \epe_e^{k_e}
			\int \frac{\Omega}{\psipol^2}
		\left( \frac{\psipol}{\phipol} \right)^{\abs{E} - 2h}
			\ln^k \frac{\psipol^{h+1}}{\phipol^h}
			\prod_{e\in E}
			\ln^{k_e} \frac{\SP_e \psipol}{\phipol},
	\label{eq:projective-expansion}
\end{equation}
expressing each coefficient as a convergent integral of a rational linear combination of products of logarithms.
Clearly this integrand can develop singularities on the coordinate hypercube $B_E$ given by the union of the faces $\SP_e = 0$ and $\SP_e \rightarrow\infty$ for any edge $e\in E$.
However, examining the denominators and arguments of logarithms we find additional singularities in the vanishing locus of any of the polynomials in
\begin{equation}
	S_{\emptyset} \defas \set{\psipol,\phipol}.
	\label{eq:initial-singularities}
\end{equation}
After integrating out a set $I \subset E$ of edges, the algorithm of \cite{Brown:TwoPoint} produces a polylogarithm with singularities contained in $B_{E\setminus I}$ and the vanishing locus $\bigcup_{f\in S_I} \set{f=0}$ of some irreducible polynomials $S_I \subset \Q[\setexp{\SP_e}{e \notin I}]$.
Studying the geometry of these \emph{Landau varieties} and obtaining small upper bounds on them is key to
\begin{enumerate}
	\item understand whether \eqref{eq:projective-expansion} can be integrated using hyperlogarithms at all,
	\item constrain the possible periods in the final result and
	\item efficient practical computation, since the size of the algebra of hyperlogarithms employed grows very sensitively with the size of $S_I$.
\end{enumerate}
We recommend very much the comprehensive in-depth discussions of \cite{Brown:PeriodsFeynmanIntegrals} (containing a wealth of insights into the underlying geometry) and recall the polynomial reduction developed therein. It keeps track of \emph{compatibilities} $C_I \subset \binom{S_I}{2}$ between the polynomials $S_I$, constituting the edges of the \emph{compatibility graph} $(S_I,C_I)$.
Starting with the complete graph $C_{\emptyset} \defas \set{\set{\psipol,\phipol}}$, for any $I \subsetneq E$ and $e \in E\setminus I$ we define $S_{I,e}$ as the set of irreducible factors (bar any monomials $\SP_{e'}$) of the polynomials
\begin{equation}
	\setexp{\resultant{0}{f}{e}, \resultant{\infty}{f}{e}, \discriminant{f}{e}}{f \in S_I} 
	\cup
	\setexp{\resultant{f}{g}{e}}{\set{f, g} \in C_I}.
	\label{eq:cg-single-reduction}
\end{equation}
Here the discriminant is $\discriminant{f}{e} \defas \resultant{f}{\partial_{\SP_e} f}{e}$ for the resultant $\resultant{f}{g}{e}$ of $f$ and $g$ with respect to $\SP_{e}$. By convention, $\resultant{0}{f}{e}$ and $\resultant{\infty}{f}{e}$ denote the constant and the leading coefficients of $f$ (with respect to the variable $\SP_{e}$).

The compatibilities $C_{I,e}$ are defined between all pairs of irreducible factors of $\resultant{f_1}{f_2}{e} \cdot \resultant{f_2}{f_3}{e}$ where $f_1,f_2,f_3 \in \set{0,\infty} \cup S_I$ and also between the irreducible factors of $\discriminant{f}{e}$ for any $f \in S_I$.
Finally we intersect over all orders of integration in
\begin{equation}
	S_{I}	\defas
		\bigcap_{e \in I} S_{I\setminus\set{e}, e}
	\quad\text{and}\quad
	C_{I} \defas
		\bigcap_{e \in I} C_{I\setminus\set{e}, e}
	\quad\text{whenever}\quad
	\emptyset \neq I \subset E.
	\label{eq:cg-recursion}
\end{equation}
\begin{definition}%
	\label{def:linearly-reducible}%
	We call $G$ \emph{linearly reducible} if there exists an ordering $\set{e_1,\ldots,e_{\abs{E}}} = E$ of its edges such that for every $1\leq i < \abs{E}$, all $f\in S_{\set{e_1,\ldots,e_{i-1}}}$ are linear in $\SP_{e_i}$.
\end{definition}
Only when all $f \in S_I$ are linear in some $\SP_e \in E \setminus I$ we can integrate out $\SP_e$ in the next step as described in \cite{Brown:TwoPoint}, so for linearly reducible $G$ we can compute its $\varepsilon$-expansion from \eqref{eq:projective-expansion} by integrating out the Schwinger parameters in any order fulfilling definition~\ref{def:linearly-reducible} and setting $\SP_{e_{\abs{E}}} = 1$.

If $e_0$ denotes the edge connecting the external legs of the propagator $G$ in the glued vacuum graph $\CloseProp{G}$, then by $\psipol_{\CloseProp{G}} = \phipol_G + \SP_{e_0} \psipol_G$ we see how linear reducibility of $G$ is equivalent to that of $\CloseProp{G}$ for which we replace \eqref{eq:initial-singularities} with $S_{\emptyset} \defas \set{\psipol_{\CloseProp{G}}}$ and $C_{\emptyset} \defas \emptyset$.

\begin{example}\label{ex:twoloops}
The simpler \emph{Fubini reduction algorithm} was used in \cite{Brown:TwoPoint} to prove linear reducibility of the wheel with three spokes $W_3$ and therefore also of the two-loop propagator
\begin{equation}
	F \defas \Graph{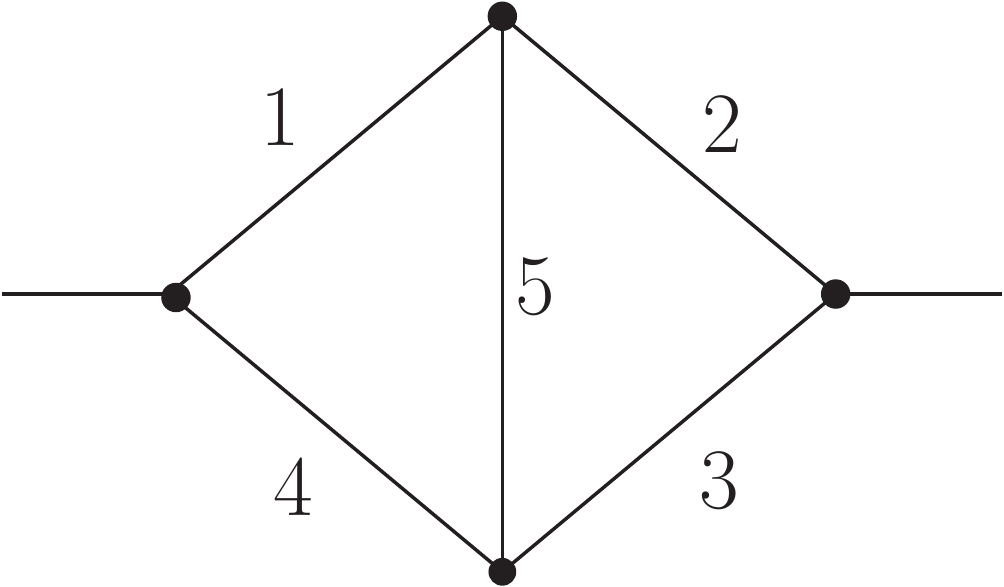}
	\quad\text{as it glues to}\quad
	\CloseProp{F} 
	= W_3
	= K_4
	= \Graph{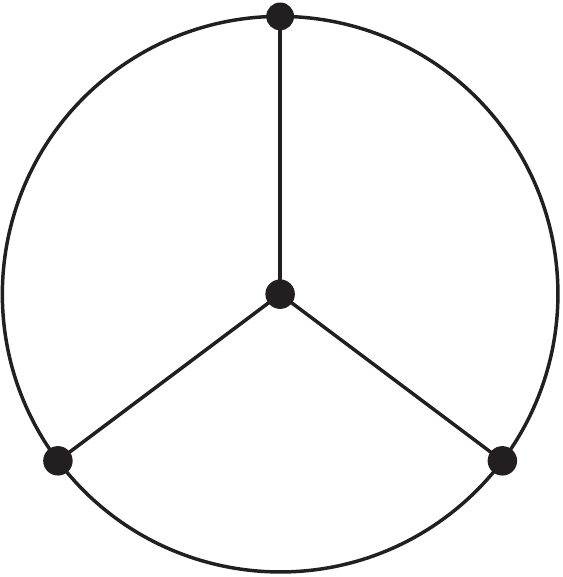}.
	\label{eq:twoloops}
\end{equation}
We refer to \cite{Grozin:TwoLoop} for a review of results and techniques to study $F$, which apart from \eqref{eq:oneloop} has so far been the only non-trivial $p$-integral computable to arbitrary order \cite{BierenbaumWeinzierl:TwoPoint} in $\varepsilon$.

To test our implementation we used it to calculate $\Phi_F$ for general $\epe_1,\ldots,\epe_5$ including the $\varepsilon^6$ term and found agreement with the results given in \cite{BierenbaumWeinzierl:TwoPoint}.
\end{example}
\begin{table}
	\centering
	\begin{tabular}{rcccccccc}
		\toprule
			Coefficient	& $\varepsilon^0$ 	 & $\varepsilon^1$	& $\varepsilon^2$	&	$\varepsilon^3$	&	$\varepsilon^4$	& $\varepsilon^5$	&	$\varepsilon^6$ & $\varepsilon^7$ \\
		\midrule
			Time in s			& 0.4	&	0.9 & 3	&	15 &	113 &	573	& 8923 & 88791 \\
			Memory in MB	&	14	& 25 &	71 & 315 & 381	&	 461 & 1225 & 2499 \\
		\bottomrule
	\end{tabular}
	\caption{Time and memory requirements for the computation of the $\varepsilon$-expansion of the two-loop propagator $F$ of example~\ref{ex:twoloops} in the case $\epe_1=\ldots=\epe_5 = 0$.}
	\label{tab:2loop-performance}
\end{table}

\section{Propagators with three loops}%
\label{sec:3loops}

Figures~\ref{fig:3loops} and \ref{fig:3loops-completions} depict the five distinct propagators with three loops as well as their glueings. 

\begin{figure}
	\begin{equation*}
		F' \defas \Graph{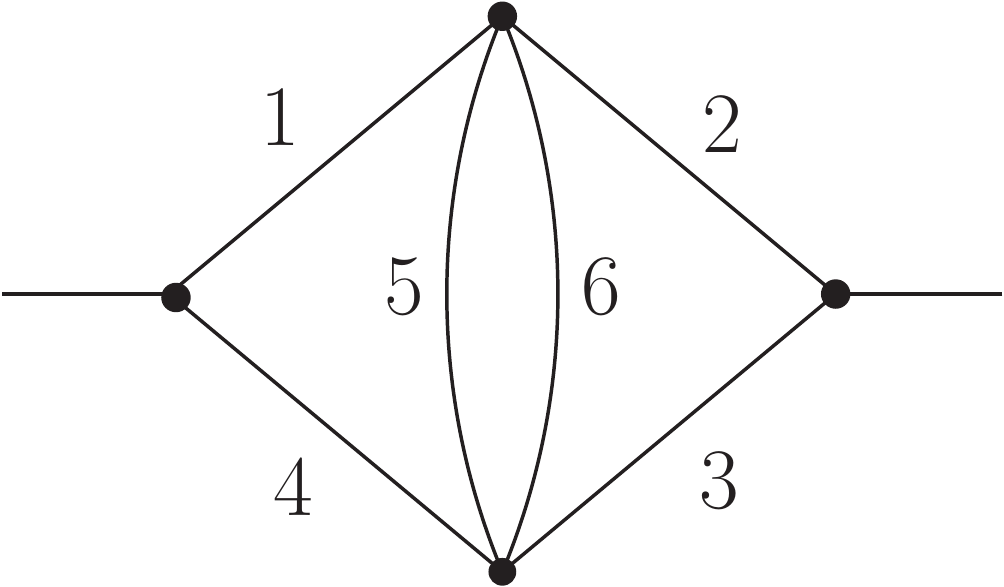}
		\qquad
		M_{4,2} \defas \Graph{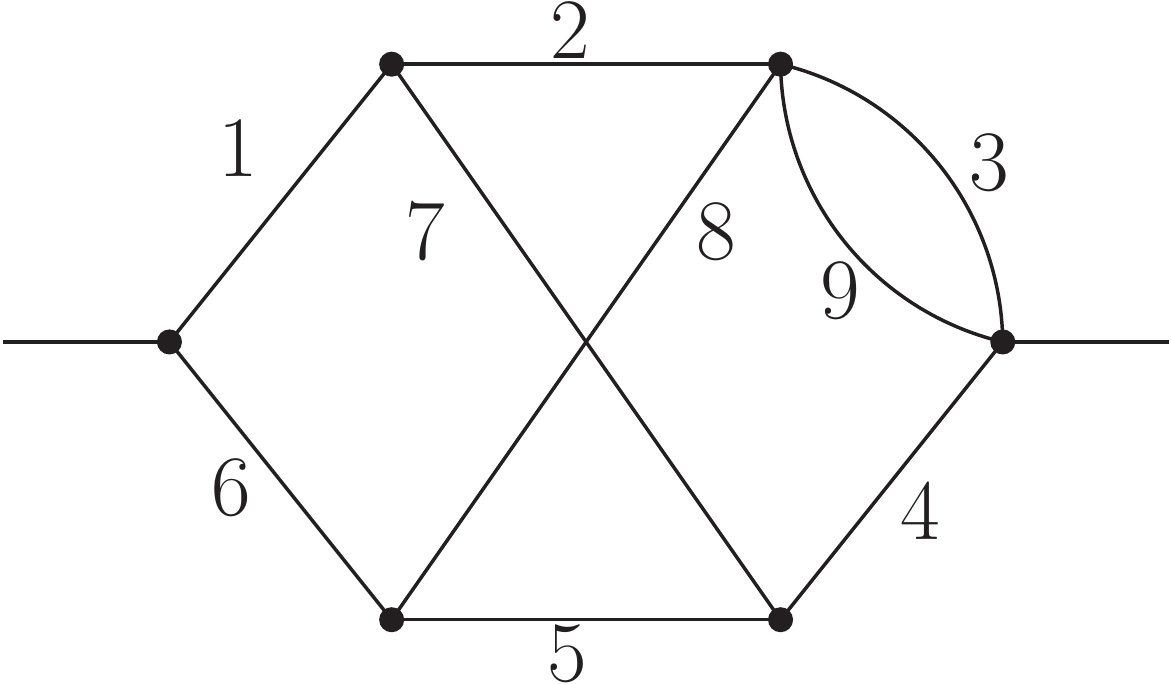}
	\end{equation*}
	\caption{%
	Graphs with one-scale subdivergences like these two factorize into smaller graphs as \eqref{eq:insertions-F'} and \eqref{eq:insertions-M42}, wherefore they do not necessitate a separate integration for evaluation.
	Other such \emph{reducible} graphs are shown in figure~\ref{fig:4loops-reducibles}.
	}
	\label{fig:insertions}
\end{figure}

Note that we consider only graphs $G$ free of any \emph{one-scale} subgraphs $\gamma$, by which we mean a connected subgraph of at least two edges that touches the edges $E\setminus E_{\gamma}$ of its complement in at most two vertices.
This is no restriction as in such a case, we can integrate out $\gamma$ independently and replace it with a single edge whose corresponding propagator is raised to a suitable power. For example we can reduce the three- and four-loop graphs of figure~\ref{fig:insertions} to
\begin{align}
	\Phi_{F'}
	&= G(\ep_5,\ep_6) \Phi_F\left(\ep_1,\ep_2,\ep_3,\ep_4,\ep_5+\ep_6-\tfrac{\D}{2}\right)
	\label{eq:insertions-F'}
	\quad\text{and}\\
	\Phi_{M_{4,2}}
	&= G(\ep_3,\ep_9) \Phi_N\left( \ep_1,\ep_2,\ep_3 + \ep_9-\tfrac{\D}{2}, \ep_4,\ep_5,\ep_6,\ep_7,\ep_8 \right).
	\label{eq:insertions-M42}
\end{align}
Even though we decided to present results only for propagator powers $\ep_e = 1 + \varepsilon\epe_e$ near unity, with our method we can equally well compute expansions around any integer powers $\ep_e = \restrict{\ep_e}{\varepsilon=0} + \varepsilon \epe_e$ where $\restrict{\ep_e}{\varepsilon=0} \in \Z$.

\subsection{The finite planar graphs $L$, $Q$ and $V$}
\label{sec:LQV}
As mentioned before, theorem~\ref{theorem:3loops} was proved in \cite{Brown:TwoPoint} for the propagators $Q$ and $V$ which are cuts of $W_4$.  
When we set sufficiently many $\epe_e = 0$, the triangles present in these graphs allow for a reduction to the two-loop graph $F$ by the standard methods of \cite{ChetyrkinTkachov:IBP}:
\begin{align}
	&\restrict{\Phi_V}{\epe_1=\epe_2=\epe_3=0}
	= \frac{1}{\varepsilon(2+\epe_6+\epe_7)} \Big\{
			\ep_6 G(\ep_5,1+\ep_6) \Phi_F(1,1,\ep_4,1+\varepsilon(1+\epe_5+\epe_6),\ep_7) \nonumber\\
			&\qquad		+ \ep_7 G(\ep_4,1+\ep_7) \Phi_F(1,1,1+\varepsilon(1+\epe_4+\epe_7),\ep_5,\ep_6)
			\label{eq:V-triangle}\\
	&\qquad		- \big[
		\ep_6 G(1+\ep_6,\ep_7)
		+ \ep_7 G(\ep_6,1+\ep_7)
		\big] \Phi_F(1,1,\ep_4,\ep_5,1+\varepsilon(1+\epe_6+\epe_7))
	\Big\},
	\nonumber\\
	&\restrict{\Phi_Q}{\epe_2=\epe_6=\epe_7=0}
	=\frac{1}{\varepsilon(2+\epe_1+\epe_5)} \Big\{
		\ep_1 G(1+\ep_1,1+\varepsilon(2+\epe_3+\epe_4+\epe_5)) \Phi_F(\ep_5,1,\ep_3,\ep_4,1)
		\nonumber\\
		&\qquad +\ep_5 G(\ep_4,1+\ep_5) \Phi_F(\ep_1,1,\ep_3,1+\varepsilon(1+\epe_4+\epe_5),1) 
		\label{eq:Q-triangle}\\
		&\qquad -\big[
			\ep_1 G(1+\ep_1,\ep_5)
			+ \ep_5 G(\ep_1,1+\ep_5)
		\big] \Phi_F(1+\varepsilon(1+\epe_1+\epe_5),1,\ep_3,\ep_4,1)
	\Big\}.
	\nonumber
\end{align}
We tested these against our computation including $\varepsilon^3$ contributions and found agreement. Further we compared with the $\bigo{\varepsilon^2}$-result for arbitrary $\epe_e$ given in \cite{Kazakov:Uniqueness}.
So while
\begin{align}
	&\frac{\Phi_V(1,\ldots,1)}{G_0^3 (1-2\varepsilon)^2}
	=	\label{eq:V-0000000}
\input{Expansions/V-0000000}
	\nonumber\\
	&\frac{\Phi_Q(1,\ldots,1)}{G_0^3 (1-2\varepsilon)^2}
	= \label{eq:Q-0000000}
\input{Expansions/Q-0000000}
	\nonumber
\end{align}
are trivial in that they follow from \eqref{eq:V-triangle} and \eqref{eq:Q-triangle}, we also computed
\begin{align}
	&\frac{\Phi_V(1+\varepsilon,\ldots,1+\varepsilon)}{G_0^3 (1-2\varepsilon)^2}
	= \label{eq:V-1111111} 
\input{Expansions/V-1111111},
	\nonumber\\
	&\frac{\Phi_Q(1+\varepsilon,\ldots,1+\varepsilon)}{G_0^3 (1-2\varepsilon)^2}
	= \label{eq:Q-1111111}
\input{Expansions/Q-1111111}.
	\nonumber
\end{align}
The complete result including $\varepsilon^4$-contributions for arbitrary $\epe_e$ is available at \cite{Panzer:MasslessPropagatorsData} and too huge to be printed here. Each {\MZV} of weight $5 + k \leq 9$ is multiplied with a polynomial in $\Q[\epe_1,\ldots,\epe_7]$ of degree $\leq k$ as illustrated in
\begin{align}
	\frac{\Phi_Q}{G_0^3 (1-2\varepsilon)^2}
	&=	%
\input{Expansions/Q-full}
	\label{eq:Q-full}\\
	p_1 &= %
\input{Expansions/Q-p_1} \\
	p_2 &= %
\input{Expansions/Q-p_2}
	\label{eq:Q-polynomials}
\end{align}
where we abbreviate $\epe_{e_1\cdots e_r} \defas \epe_{e_1} + \cdots + \epe_{e_r}$. Note how in this case $\zeta_{3,5}$ only occurs for $\epe_{2367} \neq 0$, in contrast to the propagator $V$ where
\begin{align}
	\frac{\Phi_V}{G_0^3 (1-2\varepsilon)^2}
	&= %
\input{Expansions/V-full}	\label{eq:V-full}\\
	p_1 &= %
\input{Expansions/V-p_1} \\
	p_2 &= %
\input{Expansions/V-p_2}. %
	\label{eq:V-polynomials}
\end{align}

The triangular prism $Y_3$ has vertex-width 3, implying theorem~\ref{theorem:3loops} in case of the convergent propagator $L$ through theorems 2 and 118 of \cite{Brown:PeriodsFeynmanIntegrals}. 
In the case when all $\epe_e=0$ the triangle rule can again be used to deduce
\begin{align}
	&\frac{\Phi_L(1,\ldots,1)}{G_0^3(1-2\varepsilon)^2}
	= %
\input{Expansions/L_00000000}
	\label{eq:L_00000000}
\end{align}
which we used as a check, but we computed for arbitrary $\epe_e$ and can for example give
\begin{align}
	&\frac{\Phi_L(1+\varepsilon,\ldots,1+\varepsilon)}{G_0^3(1-2\varepsilon)^2}
	= %
\input{Expansions/L_11111111}.
	\label{eq:L_11111111}
\end{align}
Our full $\varepsilon^4$-result is available at \cite{Panzer:MasslessPropagatorsData} and takes the form
\begin{align}
	&\frac{\Phi_L}{G_0^3(1-2\varepsilon)^2}
	=	\label{eq:L-full}
\input{Expansions/L-full}
	\nonumber
\end{align}
with certain polynomials $p_1,\ldots,p_5 \in \Q[\epe_1,\ldots,\epe_8]$. For example we have
\begin{align}
	p_1 & = %
\input{Expansions/L-p_1}
	\quad\text{and}\\
	p_2 & = %
\input{Expansions/L-p_2}.
	\label{eq:L-polynomials}
\end{align}

\subsection{The infrared divergent planar propagator $M$}
\label{sec:M}
\begin{figure}
	\begin{equation*}
		M^{(1)} \defas \Graph{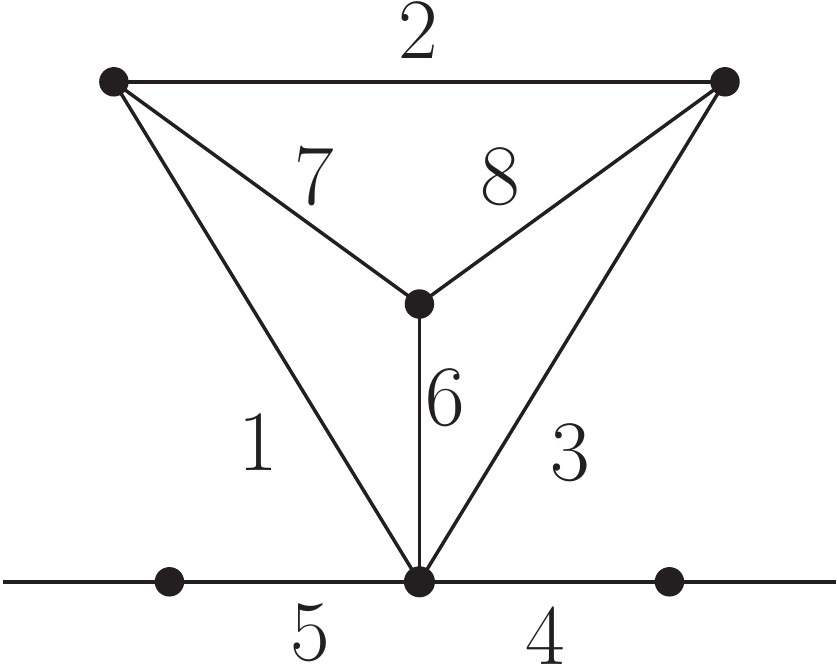}
		\qquad
		M^{(2)} \defas \Graph{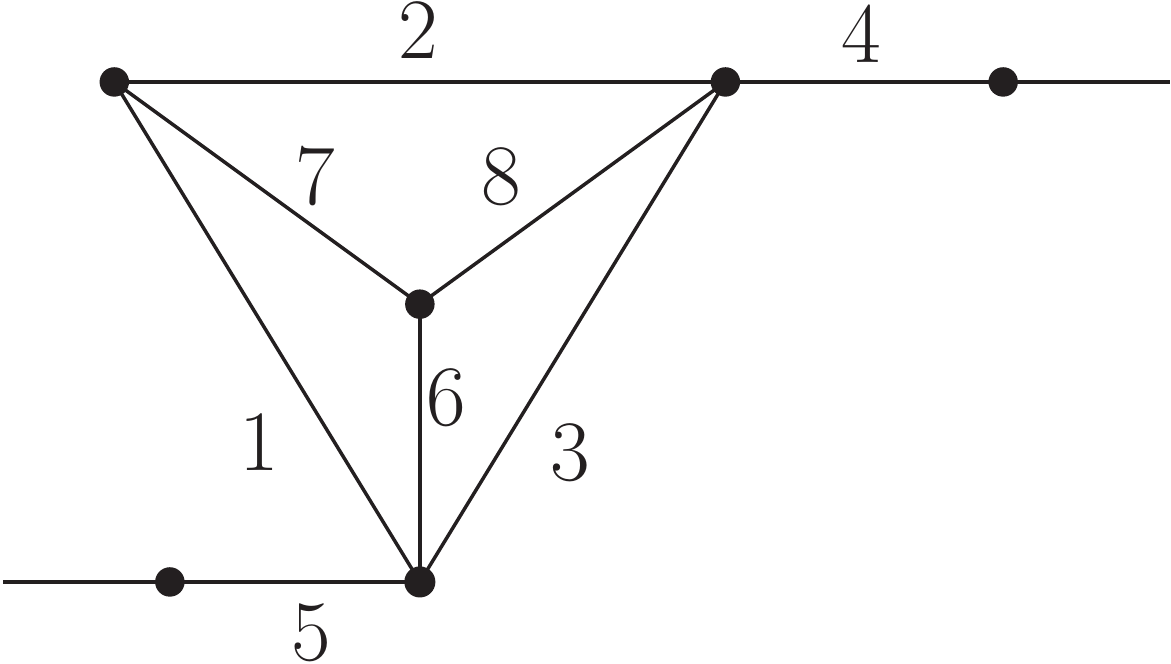}
	\end{equation*}
	\caption{These two auxiliary graphs are used in \eqref{eq:M-renormalized} as counterterms to rewrite the divergent $\Phi_M$ in terms of $\Gamma$-functions and convergent parametric integrals.}
	\label{fig:M-counterterms}
\end{figure}

An infrared subdivergence is present in the graph $M$ such that \eqref{eq:FI-parametric-projective} is divergent and can not be integrated directly by our method. We therefore renormalize by adding suitable counterterms, explicitly we rewrite
\begin{equation}
	\Phi_M
	= \frac{\Gamma(\sdc)}{\prod_{e\in E} \Gamma(\ep_e)} \left[
			I_M - I_{M^{(1)}} + I_{M^{(2)}}
		\right]
		+ \Phi_{M^{(1)}}
		- \Phi_{M^{(2)}}
	\label{eq:M-renormalized}
\end{equation}
with propagators $M^{(1)}$ and $M^{(2)}$ shown in figure~\ref{fig:M-counterterms}. Those are chosen such that $M$ and $M^{(1)}$ share the same logarithmic infrared divergence at $\SP_4, \SP_5 \rightarrow 0$ while the logarithmic ultraviolet divergence $\SP_1,\SP_2,\SP_3,\SP_6,\SP_7,\SP_8 \rightarrow 0$ is common to both $M^{(1)}$ and $M^{(2)}$.
In particular $I_M - I_{M^{(1)}} + I_{M^{(2)}}$ is a finite (convergent) parametric integral which we can compute using hyperlogarithms.
Now we only need to add back the counterterms in \eqref{eq:M-renormalized} which is simple since the tadpole forces $\Phi_{M^{(1)}} = 0$ in dimensional (also in analytic) regularization and the subdivergence of $M^{(2)}$ is one-scale such that
\begin{equation}
	\Phi_{M^{(2)}}
	= G \Big(
				\ep_3,
				1+\varepsilon(2+\epe_1+\epe_2+\epe_6+\epe_7+\epe_8)
		\Big)
		\Phi_F(\ep_1,\ep_2,\ep_8,\ep_6,\ep_7).
	\label{eq:Mct2}
\end{equation}
Again we calculated to $\varepsilon^4$ for arbitrary $\epe_e$ and successfully compared the simple case
\begin{align}
	&\frac{\Phi_M(1,\ldots,1)}{G_0^3 (1-2\varepsilon)^2} \cdot 3(1+2\varepsilon)
	= %
\input{Expansions/M-00000000}
	\label{eq:M-00000000}
\end{align}
to the formula in terms of $G$- and $\Phi_F$-functions obtained by applying the triangle rule:
\begin{align}
	&\frac{1+2\varepsilon}{G_0}
		\Phi_M(1,\ldots,1)
		= 
		-3 \frac{1-2\varepsilon}{\varepsilon} \Phi_F(1,1,1,1,1+\varepsilon)
	\label{eq:M-triangle}\\&\quad
		+ \frac{\Gamma^2(-\varepsilon)}{\varepsilon^2}
			\left[
			 	\frac{
						\Gamma( 1+2\varepsilon)
						\Gamma( -\varepsilon)
				}{
						\Gamma( -3\varepsilon)
				}
				- \frac{
						\Gamma(1+3 \varepsilon)
						\Gamma(-3\varepsilon)
						\Gamma(-\varepsilon)
						\Gamma(\varepsilon)
				}{
						\Gamma(-2\varepsilon)
						\Gamma(-4\varepsilon)
						\Gamma(2\varepsilon)
				}
				+ \frac{
							7
							\Gamma(-2\varepsilon)
							\Gamma(3\varepsilon)
				}{
							2
							\Gamma(-4\varepsilon)
							\Gamma(\varepsilon)
				}
			\right].
	\nonumber
\end{align}
An example for non-vanishing $\epe_e$ is given by
\begin{align}
	&\frac{\Phi_M(1+\varepsilon,\ldots,1+\varepsilon)}{G_0^3 (1-2\varepsilon)^2} \cdot 9(1+8\varepsilon)
	= %
\input{Expansions/M-11111111},
	\label{eq:M-11111111}
\end{align}
with the full $\varepsilon^4$-result available at \cite{Panzer:MasslessPropagatorsData}. The first terms read
\begin{align}
	&\frac{\Phi_M\cdot (3+\epe_{123678}) (1+\varepsilon[2 + \epe_{123678}])}{G_0^3 (1-2\varepsilon)^2}
	= %
\input{Expansions/M-full}
	\label{eq:M-full}
\end{align}
with the polynomial $p_1$ given by
\begin{align}
	p_1 &= %
\input{Expansions/M-p_1}.
	\label{eq:M-polynomials}
\end{align}
The rational prefactor on the left-hand side of \eqref{eq:M-full} is chosen such that the $\varepsilon^k$-period on the right-hand side is of homogeneous weight $4+k$.

\subsection{The non-planar convergent graph $N$}
\label{sec:N}

The most interesting propagator $N$ does not feature vertex-width three, but \cite{Brown:PeriodsFeynmanIntegrals} already observed its linear reducibility by explicit computation using the compatibility graph method of section~\ref{sec:reduction}. We like to point out that the simple Fubini reduction algorithm of \cite{Brown:TwoPoint} does not suffice here and can only show that the coefficients of $N$ are expressible as multiple polylogarithms evaluated at sixth roots of unity.

We chose the edge sequence $(e_1,\ldots,e_8) = (1, 6, 3, 4, 5, 2, 7, 8)$ for integration and found that actually all polynomials $S_{\set{e_1,\ldots,e_{i-1}}}$ are linear in all variables, not just $\SP_{e_i}$. By
\begin{align}
	S_{\set{1,6,3,4,5,2}}
	= \set{\SP_7+\SP_8, \SP_7 - \SP_8}
	\label{}
\end{align}
the algorithm produces hyperlogarithms with singularities in $\set{-1, 0, 1}$ (recall that $\SP_7$ is the last integration and we set $\SP_8 = 1$), yielding alternating Euler sums. We computed the full expansion to $\varepsilon^4$ for arbitrary $\epe_e$ but found that all such sums in the periods obtained actually combined to {\MZV}.

Our result for the special case $\epe_e = 0$ is \eqref{eq:N-expansion} and agrees with the numerical investigation of \cite{LeeSmirnov:EasyWay}.
For arbitrary $\epe_e$ we get
\begin{align}
	\frac{\Phi_N}{G_0^3(1-2\varepsilon)^2}
	&= %
\input{Expansions/N-full} \\
	p_1 &= %
\input{Expansions/N-p_1} \\
	p_2 &= %
\input{Expansions/N-p_2}
	\label{eq:N-full}
\end{align}
while the full result including $\varepsilon^4$ can be obtained from \cite{Panzer:MasslessPropagatorsData}. For example,
\begin{align}
	&\frac{\Phi_N(1+\varepsilon,\ldots,1+\varepsilon)}{G_0^3(1-2\varepsilon)^2}
	= %
\input{Expansions/N-11111111}.
	\label{eq:N-11111111}
\end{align}
\begin{table}
	\centering
	\begin{tabular}{rrrrrr}
		\toprule
			Coefficient	& $\varepsilon^0$ 	 & $\varepsilon^1$	& $\varepsilon^2$	&	$\varepsilon^3$	&	$\varepsilon^4$	\\
		\midrule
		Time		& 1s & 11s	&	400s &	\hide{14665 s} 4h &	\hide{358064 s} 100h	\\
			Memory in MB	& 33 &	398 & 847 & 1625	&	 11119\\
		\bottomrule
	\end{tabular}
	\caption{Time and memory requirements for the computation of the $\varepsilon$-expansion \eqref{eq:N-expansion} of $N$ without insertions, that is $\epe_1=\ldots=\epe_8 = 0$.}
	\label{tab:N-performance}
\end{table}

\subsection{Checks and symmetries}
\label{sec:3loop-symmetries}

The automorphisms $\Aut(G) \subseteq S\left(E_{G}\right)$ of a propagator constitute a subgroup of permutations of its edges and the integral $\Phi_G$ is invariant under the action of this group. For all considered graphs we explicitly verified that our results do indeed obey this symmetry.

Note how this property serves a highly non-trivial check of our implementation: Since we fix an order $e_1,\ldots,e_{\abs{E}}$ of integration this symmetry is not manifest in the algorithm at all and indeed the intermediate results differ considerably when we choose a different order. 
It is only after reducing the final output to a basis of {\MZV} that the symmetry reveals itself.

\begin{example}
	The propagators $M,Q$ and $V$ possess only a single non-trivial automorphism. In disjoint cycle notation these are $(1\;3)(4\;5)(7\;8)$, $(1\;4)(2\;3)(6\;7)$ and $(1\;3)(4\;5)(6\;7)$ respectively.
\end{example}

In particular we like to stress that we did not exploit any such information on symmetries in the first place, but naively integrated all the coefficients in \eqref{eq:projective-expansion} for every single monomial in the variables $\set{\varepsilon,\epe_1,\ldots,\epe_{\abs{E}}}$.

The glueing process provides many more relations, in particular for the propagators $V$ and $Q$. If in general we let $e_0 \in E_{\CloseProp{G}}$ denote the glued edge, then integration of $\SP_{e_0}$ proves
\begin{equation}
	I_{G} 
	\urel{\eqref{eq:FI-parametric-projective}}
	\frac{
		\Gamma\left( \tfrac{\D}{2} \right)
		I_{\CloseProp{G}}
	}{
		\Gamma(\sdc_{G}) \Gamma(\ep_{e_0})
	}
	\quad\text{and}\quad
	\Phi_{G} 
	\urel{\eqref{eq:FI-parametric-affine}}
	\frac{
		\Gamma\left( \tfrac{\D}{2} \right)
		I_{\CloseProp{G}}
	}{
		\prod_{e\in E_{\CloseProp{G}}} \Gamma(\ep_{e})
	}
	\quad\text{when we set}\quad
	\ep_{e_0}
	\defas \tfrac{\D}{2} - \sdc_{G}.
	\label{eq:glueing}
\end{equation}
Hence from $\Phi_G$ we can compute $I_{\CloseProp{G}}$ which we view as a function of $\ep_{e_0}, \ldots, \ep_{e_{\abs{E}}}$ only and as such enjoys the full symmetry $\Aut\big( \CloseProp{G} \big)$ of the glued graph. 
Since $\sdc = 1 + \varepsilon(3 + \epe_{1234567})$ for $G\in\set{V,Q}$ and therefore $a_0 = 1 -\varepsilon(4 + \epe_{1234567})$, such that we expand all propagator powers $\ep_e$ of $\CloseProp{G}$ around unity, these symmetries push down to the $\varepsilon$-expansions $\Phi_G$ viewed as series in $\ep_0-1,\ldots,\ep_7-1$.

\begin{figure}
	\begin{equation*}
		W_4 = \Graph{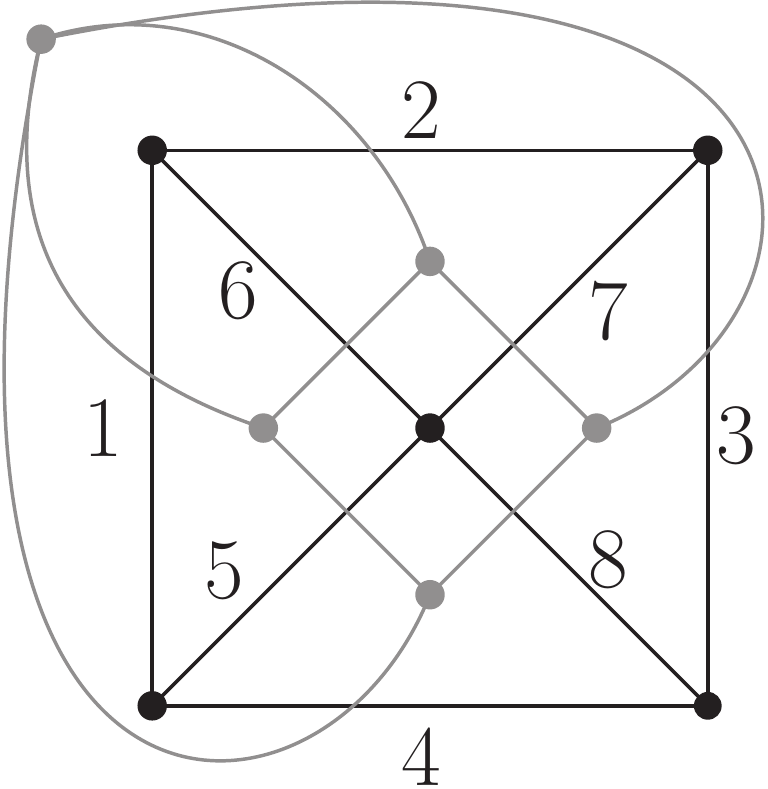}
		\qquad\text{has the planar dual}\qquad
		\Fourier{W_4} = \Graph{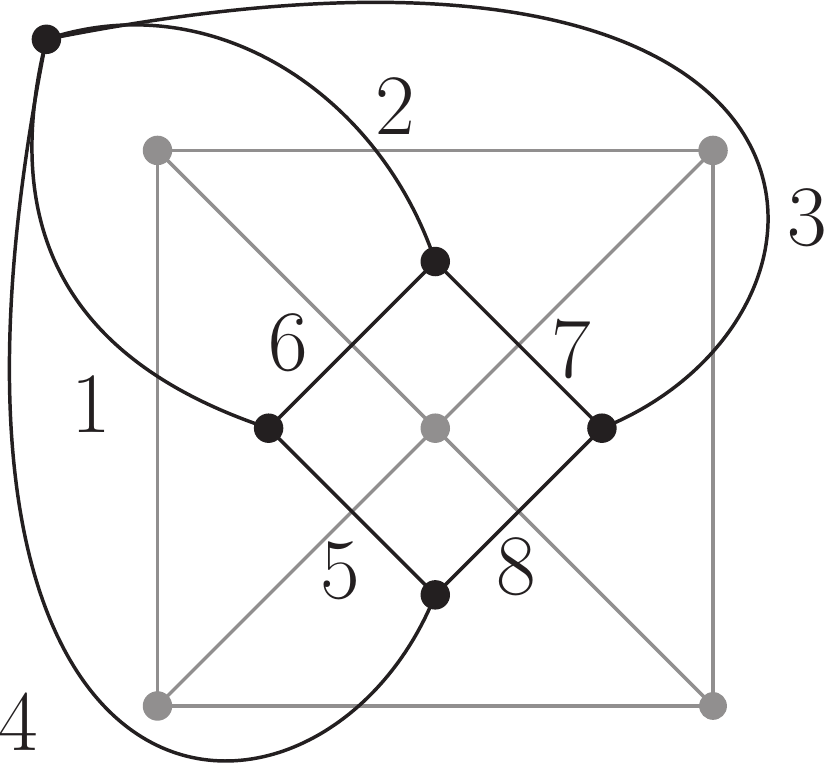}
		= \sigma\left( W_4 \right)
	\end{equation*}
	\caption{%
		The planar dual of $W_4$ is isomorphic to $W_4$, for example under the permutation $\sigma = (5\;6\;7\;8\;4\;1\;2\;3): E_{W_4} \rightarrow E_{\Fourier{W_4}}$. This implies the relation given in \eqref{eq:fourier}.
	}
	\label{fig:W4-duality}
\end{figure}

Altogether both $Q$ and $V$ are equivalent and transform into each other by
\begin{align}
	I_{V}(\ep_1,\ldots,\ep_7)
	&= \frac{\Gamma(2-\varepsilon)}{\Gamma(\sdc)\Gamma(\ep_0)}
	I_{W_4} \left( 
			\ep_1, \ep_2, \ep_3, \ep_0, \ep_5, \ep_6, \ep_7, \ep_4
		\right)
	\quad\text{and}\quad
	\label{eq:V-W4} \\
	I_{Q}(\ep_1,\ldots,\ep_7)
	&= \frac{\Gamma(2-\varepsilon)}{\Gamma(\sdc)\Gamma(\ep_0)}
	I_{W_4} \left( 
			\ep_2, \ep_6, \ep_7, \ep_3, \ep_0, \ep_1, \ep_5, \ep_4
		\right)
	\label{eq:Q-W4}
\end{align}
with respect to the edge numberings of figures~\ref{fig:3loops} and \ref{fig:W4-duality}. Applying \eqref{eq:V-W4} and \eqref{eq:Q-W4} to our results (that were calculated separately for $V$ and $Q$) we verified that both give the same function $I_{W_4}$.

We also checked its $\abs{Aut(\CloseProp{G})} = \abs{\Aut(W_4)} = 8$-fold symmetry and the \emph{Fourier identity}
\begin{equation}
	I_{W_4}(\ep_1,\ldots,\ep_8)
	= I_{W_4} \left(
			\Fourier{\ep_5}, \Fourier{\ep_6}, \Fourier{\ep_7}, \Fourier{\ep_8},
			\Fourier{\ep_4}, \Fourier{\ep_1}, \Fourier{\ep_2}, \Fourier{\ep_3}
		\right)
	\quad\text{where}\quad
	\Fourier{\ep_e}
	\defas
	\tfrac{\D}{2} - \ep_e
	\label{eq:fourier}
\end{equation}
exploiting the planar self-duality $\Fourier{W_4} \isomorph W_4$ shown in figure~\ref{fig:W4-duality}. Note that \eqref{eq:fourier} is just a consequence of transforming the Schwinger parameters as $\SP_e \mapsto \SP_e^{-1}$.
\begin{remark}
	There are further symmetries originating from \emph{uniqueness} \cite{Kazakov:Uniqueness}, but the analysis of the full group of symmetries of $I_{W_4}$ is a subject on its own which we hope to return to in the future. For now we refer to \cite{BarfootBroadhurst:Z2S6} where it was carried out for the tetrahedron $W_3=K_4$ and exploited in the evaluation of $\Phi_F$ to great effect.
\end{remark}
\begin{remark}
	For $N$ we expand $\ep_{e_0} = -\varepsilon (3 + \epe_{12345678})$ around zero, hence after glueing to $\CloseProp{N}=K_{3,3}$ we can apply automorphisms to obtain the expansions for $N$ when one edge power $\ep_e = \epe_e \varepsilon$ is near zero instead of one. This allows to compute \eqref{eq:insertions-M42} and we checked these results with $M_{4,1}$ and similarly $M_{4,2}$ given in \cite{LeeSmirnov:FourLoopPropagatorsWeightTwelve}.
\end{remark}

\section{Propagators with four loops}%
\label{sec:4loops}

\begin{figure}
	\begin{equation*}
		{_5 R_1} = \Graph{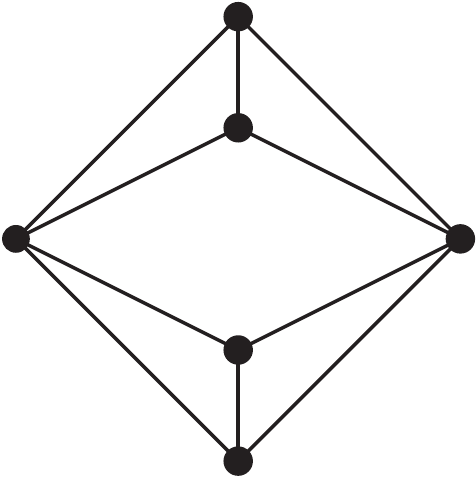}
		\quad
		{_5 R_2} = \Graph{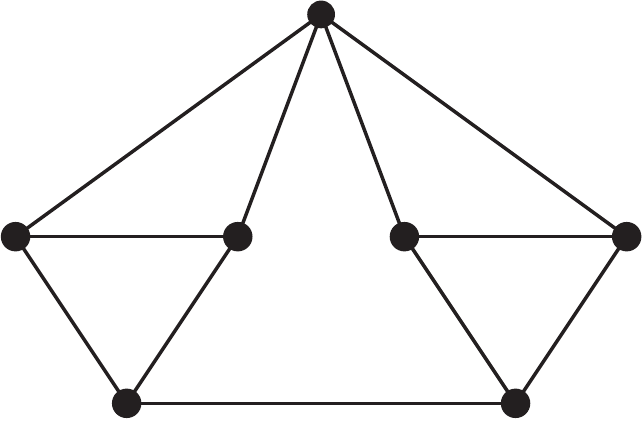}
		\quad
		{_5 R_3} = \Graph{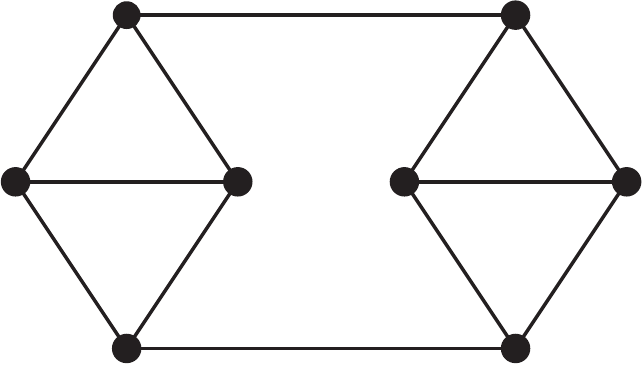}
		\quad
	\end{equation*}
	\caption{%
		These \emph{reducible} ($R$) five-loop graphs factorize into their one-scale subgraphs $F$ and thus need not be computed separately. Note that ${_5 R_1}$ is called $5 R$ in \cite{Brown:TwoPoint}.
	}
	\label{fig:4loops-reducibles}
\end{figure}

We enumerated the connected 1PI five-loop vacuum graphs without one-scale subgraphs in figure~\ref{fig:4loops-completions}. So in particular we omitted graphs with parallel edges, two-valent vertices and also those shown in figure~\ref{fig:4loops-reducibles}.

Theorem~\ref{theorem:4loops} is immediate for all the planar graphs except ${_5 P_7}$, as these feature vertex-width 3. 
We explicitly computed the polynomial reduction for ${_5 P_7}$ and the non-planar graphs and indeed verified linear reducibility in each case. 
Further we checked that the singularities of the hyperlogarithms obtained after the last integration are contained in $\set{-1,0,1}$ by considering the limits of the zeros of the polynomials in the reduction, as described in \cite{Brown:TwoPoint,Brown:PeriodsFeynmanIntegrals}. This completes the proof of theorem~\ref{theorem:4loops}.

\begin{figure}
	\begin{gather*}
		M_{3,5} = \Graph{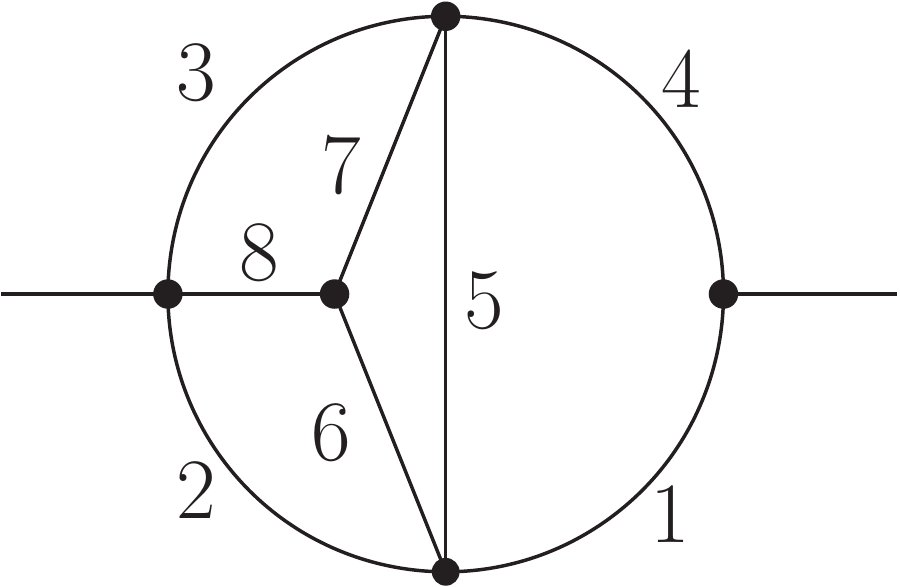}
		\quad
		M_{3,6} = \Graph{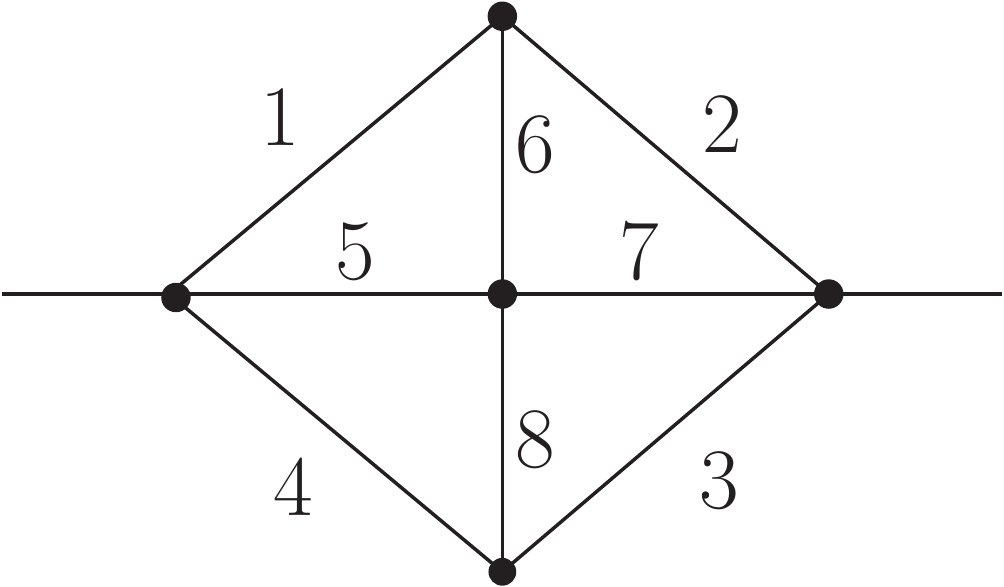}
		\quad
		M_{4,4} = \Graph{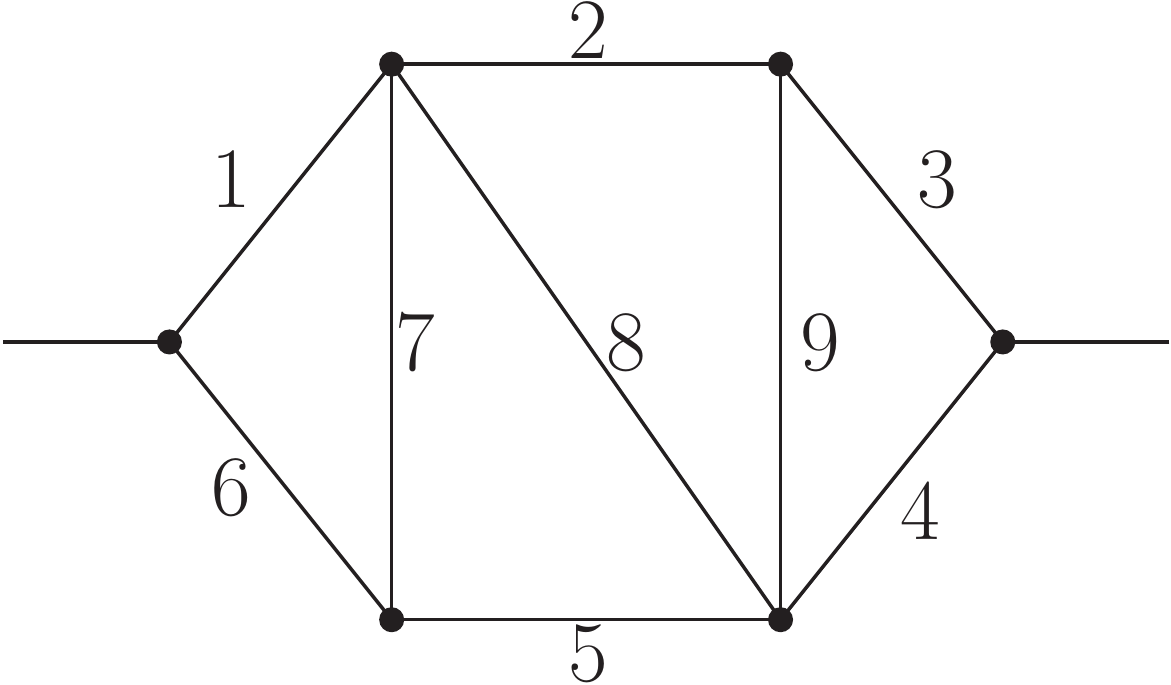}
		\\
		M_{4,5} = \Graph{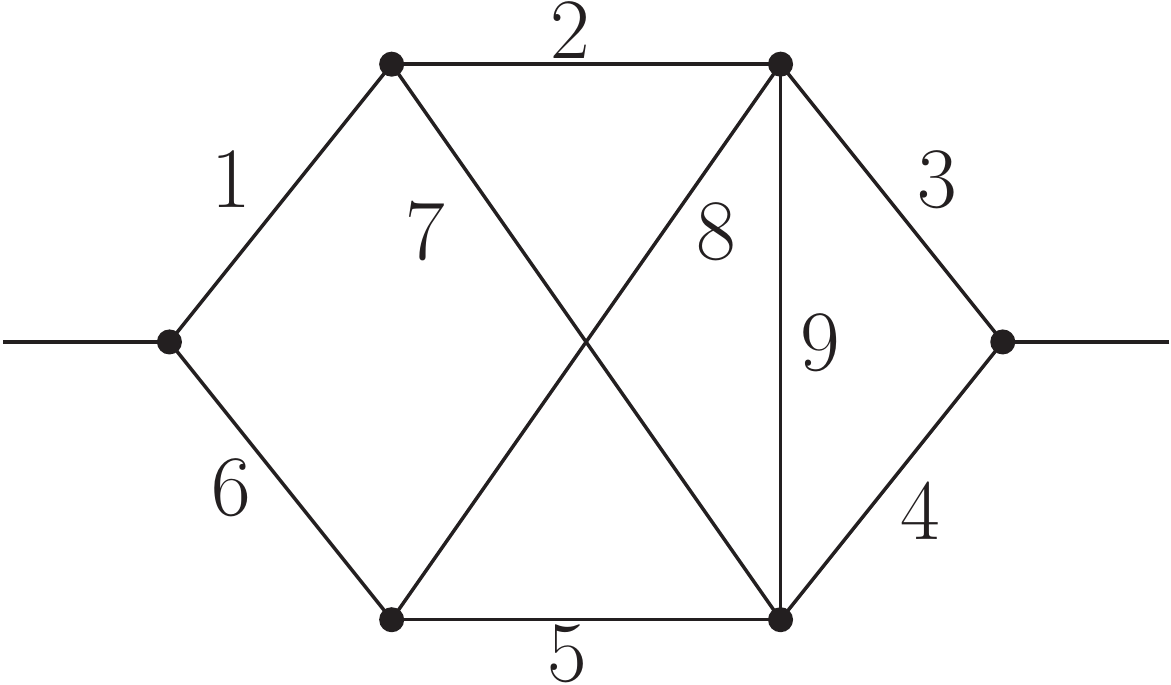}
		\quad
		M_{5,1} = \Graph{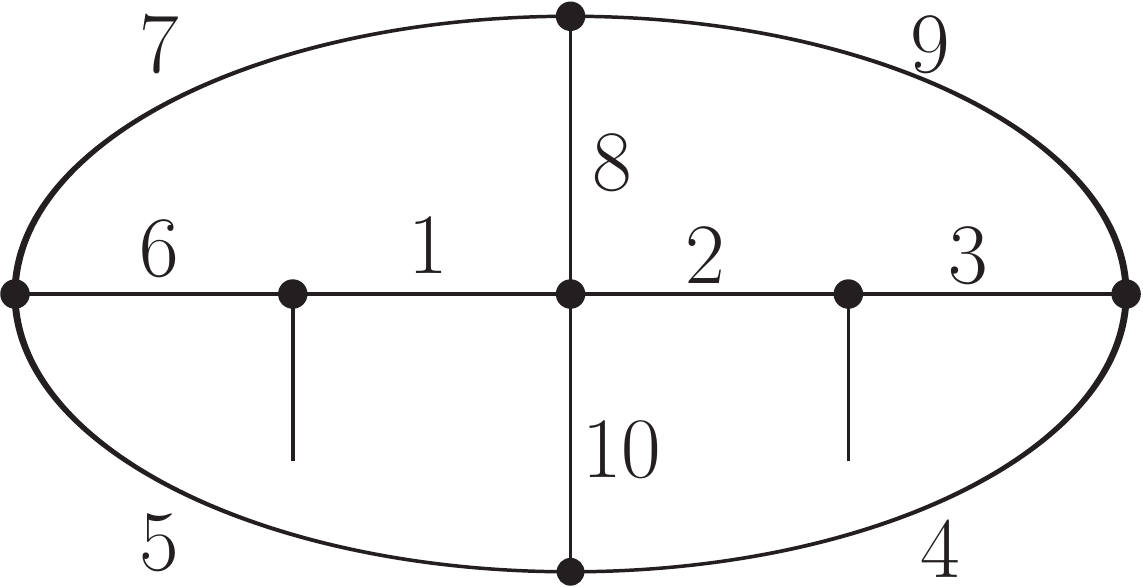}
	\end{gather*}
	\caption{%
		Some four-loop propagators we calculated that arise as master integrals as considered for example in \cite{BaikovChetyrkin:FourLoopPropagatorsAlgebraic,LeeSmirnov:FourLoopPropagatorsWeightTwelve}.
	}
	\label{fig:4loops}
\end{figure}

Given the multitude of four-loop propagators we did not calculate all of them and instead focused on the graphs shown in figure~\ref{fig:4loops} that recently enjoyed great interest as master integrals in \cite{BaikovChetyrkin:FourLoopPropagatorsAlgebraic,SmirnovTentyukov:FourLoopPropagatorsNumeric,LeeSmirnov:FourLoopPropagatorsWeightTwelve}.

In particular, a few coefficients of their expansions were obtained algebraically in \cite{BaikovChetyrkin:FourLoopPropagatorsAlgebraic} for the special case $\epe_e=0$ for all edges $e$.
We could verify these by our method and provide at least one further coefficient in the following. These were also checked to agree with the numerical calculations of \cite{LeeSmirnov:FourLoopPropagatorsWeightTwelve}.

Note however that in each case considered we provide full results for arbitrary $\epe_e$. These are necessary for calculations to higher loop numbers, for example if a one-scale subgraph is inserted into some edge like in figure~\ref{fig:insertions}.

\subsection{Propagators without subdivergences}
The logarithmically divergent, primitive wheel with four spokes $W_4=M_{3,6}$ is a cut of ${_5 P_1}$. We computed its expansion analytically including the $\varepsilon^2$ contributions (the $\varepsilon^3$-term can be obtained from our result for $L$ as we explain in section~\ref{sec:4loop-symmetries}). For example
\begin{align}
	&\frac{\Phi_{M_{3,6}}(1,\ldots,1)\cdot 4(1-5\varepsilon)}{G_0^4(1-2\varepsilon)^3}
	= %
\input{Expansions/M36-00000000}
	\quad\text{and}
	\label{eq:M36-00000000} \\
	&\frac{\Phi_{M_{3,6}}(1+\varepsilon,\ldots,1+\varepsilon)\cdot 12(1-13\varepsilon)}{G_0^4(1-2\varepsilon)^3}
	= %
\input{Expansions/M36-11111111}.
	\label{eq:M36-11111111}
\end{align}
The complete result for arbitrary $\epe_e$ has the form
\begin{align}
	&\frac{\Phi_{M_{3,6}}\cdot(1-\varepsilon[5+\epe_{12345678}])(4+\epe_{12345678})}{G_0^4(1-2\varepsilon)^3}
	= %
\input{Expansions/M36-full}
	\label{eq:M36-full}
\end{align}
for the polynomials given in
\begin{align}
	p_1 &= %
\input{Expansions/M36-p_1} \\
	p_2 &= %
\input{Expansions/M36-p_2} \\
	p_3 &= %
\input{Expansions/M36-p_3}.
	\label{eq:M36-polynomials}
\end{align}
In the case of the planar graph $M_{4,4}$, theorem~\ref{theorem:4loops} was proved in \cite{Brown:TwoPoint}. Its value $\frac{441}{8} \zeta_7$ at $\D=4$ has long been known as the period of the zig-zag graph $Z_5={_5 P_3}$. We state our result for the first three coefficients in $\varepsilon$ as
\begin{align}
	&\frac{\Phi_{M_{4,4}}}{G_0^4(1-2\varepsilon)^3}
	= %
\input{Expansions/M44-full},
	\label{eq:M44-full}
\end{align}
where we introduced the polynomials
\begin{align}
	p_1 &= 	%
\input{Expansions/M44-p_1} \\
	p_2 &= 	%
\input{Expansions/M44-p_2} \\
	p_3 &= 	%
\input{Expansions/M44-p_3}.
	\label{eq:M44-polynomials}
\end{align}
For the special case $\epe_1=\ldots=\epe_9 = 0$ we read off $p_1=p_2=p_3=0$ and therefore
\begin{align}
	\frac{\Phi_{M_{4,4}}(1,\ldots,1)}{G_0^4(1-2\varepsilon)^3}
	&= %
\input{Expansions/M44_000000000}.
	\label{eq:M44-000000000}
\hide{
	&\frac{\Phi_{M_{4,4}}(1+\varepsilon,\ldots,1+\varepsilon)}{G_0^4(1-2\varepsilon)^3}
	= %
\input{Expansions/M44_111111111}
	\label{eq:M44-111111111}
}%
\end{align}
The non-planar graph $M_{4,5}$ could not be detected as linearly reducible in \cite{Brown:TwoPoint} and was studied numerically in \cite{LeeSmirnov:EasyWay}. Note that though alternating Euler sums appear as the result of our integration, they combine to {\MZV} as far as we calculated.
We obtained
\begin{align}
	&\frac{\Phi_{M_{4,5}}\cdot(1-\varepsilon[6+p_1])}{G_0^4 (1-2\varepsilon)^3}
	=	%
\input{Expansions/M45-full}
	\label{eq:M45-full}
\end{align}
where the polynomials $p_1,p_2,p_3 \in \Q[\epe_1,\ldots,\epe_9]$ are given by
\begin{align}
	p_1 &=	%
\input{Expansions/M45-p_1}%
	\quad
	p_2 = %
\input{Expansions/M45-p_2}%
	\nonumber\\
	p_3 &= %
\input{Expansions/M45-p_3}.
	\label{eq:M45-polynomials}
\end{align}
In the special case $\epe_e=0$ we have $p_1=p_2=p_3=0$ and find
\begin{align}
	\frac{\Phi_{M_{4,5}}(1,\ldots,1)\cdot(1-6\varepsilon)}{G_0^4(1-2\varepsilon)^3}
	&= %
\input{Expansions/M45-000000000}.
	\label{eq:M45-000000000}
\end{align}

\subsection{The subdivergent propagators $M_{3,5}$ and $M_{5,1}$}
\begin{figure}
	\begin{equation*}
		M_{3,5}^{(1)} \defas \Graph{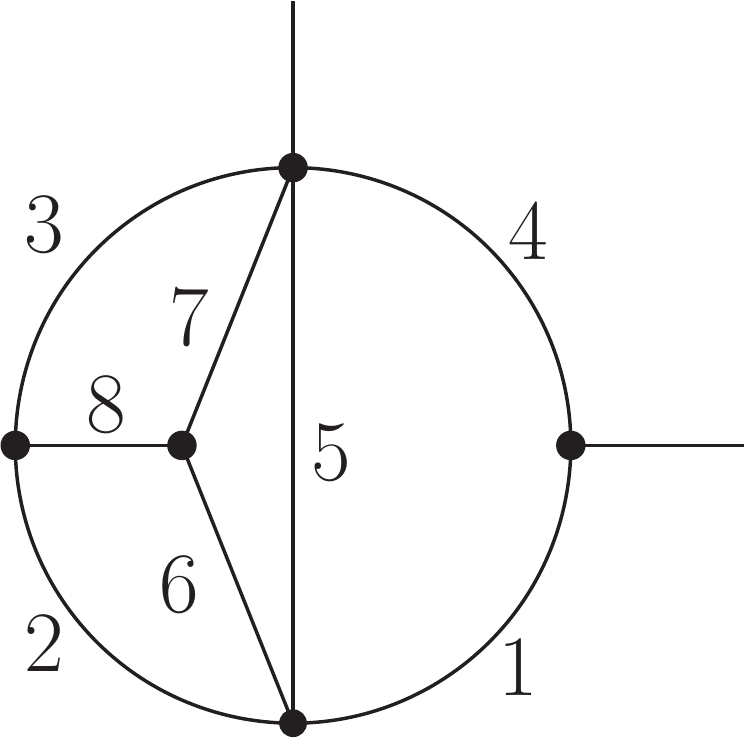}
		\qquad
		M_{5,1}^{(1)} \defas \Graph{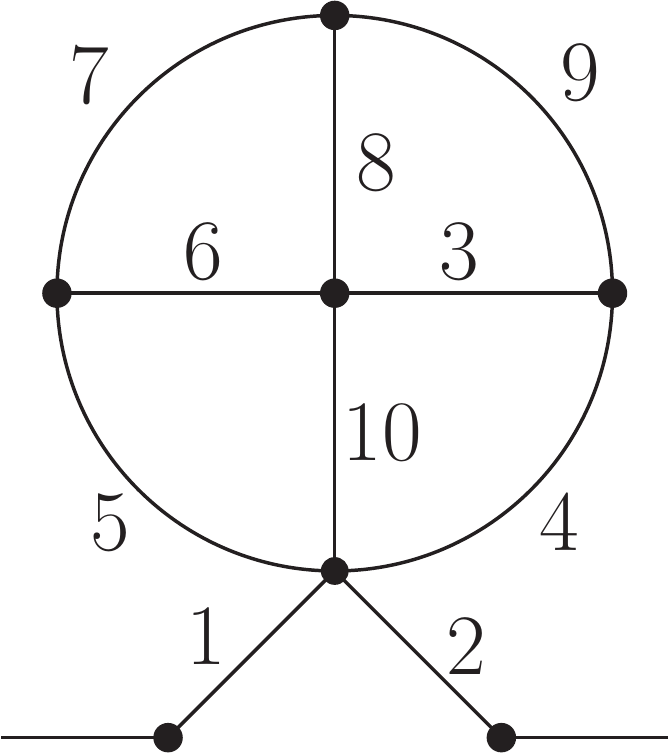}
		\qquad
		M_{5,1}^{(2)} \defas \Graph{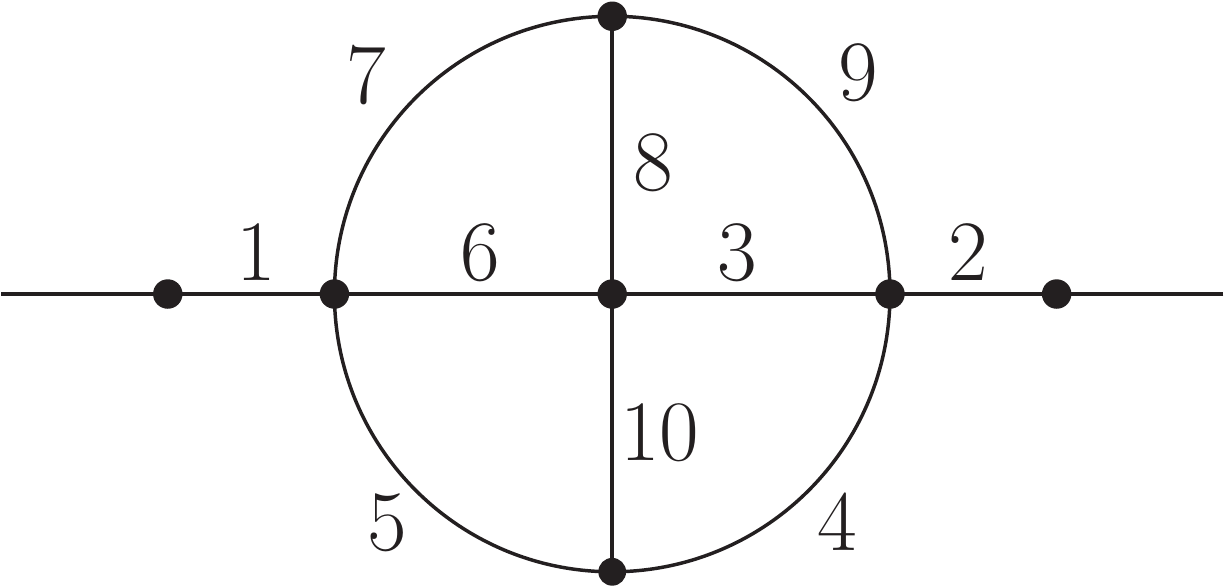}
	\end{equation*}
	\caption{The counterterm graphs employed in \eqref{eq:M35-renormalized} and \eqref{eq:M51-renormalized} to compute $M_{3,5}$ and $M_{5,1}$ which contain an ultraviolet and an infrared subdivergence, respectively.}
	\label{fig:4loop-counterterms}
\end{figure}
$M_{3,5}$ has an ultraviolet subdivergence isomorphic to $W_3$ and formed by the edges $\set{2,3,5,6,7,8}$. We subtract $M_{3,5}^{(1)}$ of figure~\ref{fig:4loop-counterterms} to cancel it and thus compute
\begin{equation}
	\Phi_{M_{3,5}}
	= \frac{\Gamma(\sdc)}{\prod_{e\in E} \Gamma(\ep_e)} \left[ 
			I_{M_{3,5}}
			- I_{M_{3,5}^{(1)}}
		\right]
		+ \Phi_{M_{3,5}^{(1)}}
	\label{eq:M35-renormalized}
\end{equation}
by integrating the convergent parametric integral $I_{M_{3,5}} - I_{M_{3,5}^{(1)}}$ and adding to it
\begin{equation}
	\Phi_{M_{3,5}^{(1)}}
	= G\left(\ep_4, 1+\varepsilon(3+\epe_{1235678}) \right)
		G\left( \ep_5, 1+\varepsilon(2+\epe_{23678}) \right)
		\Phi_F \left( \ep_7,\ep_6,\ep_2,\ep_3,\ep_8 \right).
	\label{eq:M35-ct}
\end{equation}
Note that it is always possible to find such counterterms where all subdivergences are one-scale \cite{BrownKreimer:AnglesScales}.
We calculated this expansion including $\varepsilon^3$ contributions and for example obtained
\begin{align}
	&\frac{\Phi_{M_{3,5}}(1,\ldots,1)
					\cdot 12(1-5\varepsilon)(1-4\varepsilon)
				}{G_0^4(1-2\varepsilon)^3}
	= %
\input{Expansions/M35-00000000}
	\quad\text{and}
	\label{eq:M35-00000000}	\\
	&\frac{\Phi_{M_{3,5}}(1+\varepsilon,\ldots,1+\varepsilon)
		\cdot 108(1-13\varepsilon)(1-10\varepsilon)}{G_0^4(1-2\varepsilon)^3}
	= %
\input{Expansions/M35-11111111}.
	\label{eq:M35-11111111}
\end{align}
The first few terms of our full result for arbitrary $\epe_e$ read
\begin{align}
	&\frac{\Phi_{M_{3,5}}\cdot(1-\varepsilon[5+p_3])(1-\varepsilon[4+p_4])(4+p_3)(3+p_4)}{G_0^4 (1-2\varepsilon)^3}
	=	%
\input{Expansions/M35-total}
	\label{eq:M35-full}
\end{align}
where the polynomials $p_1,\ldots,p_4 \in \Q[\epe_1,\ldots,\epe_8]$ are given by
\begin{align}
	p_1 &=	%
\input{Expansions/M35-p_1} \\
	p_2 &= %
\input{Expansions/M35-p_2} \\
	p_3 &= \epe_{12345678}
	\qquad
	p_4 =  \epe_{235678}.
	\label{eq:M35-polynomials}
\end{align}
$M_{5,1}$ features an infrared subdivergence which manifests itself in the parametric representation at $\SP_3,\ldots,\SP_{10} \rightarrow 0$. Analogously to \eqref{eq:M-renormalized} of the three-loop graph $M$ we add two counterterms shown in figure~\ref{fig:4loop-counterterms} and compute
\begin{equation}
	\Phi_{M_{5,1}}
	= \frac{\Gamma(\sdc)}{\prod_{e\in E} \Gamma(\ep_e)} \left[ 
			I_{M_{5,1}}
			- I_{M_{5,1}^{(1)}}
			+ I_{M_{5,1}^{(2)}}
		\right]
		+ \Phi_{M_{5,1}^{(1)}}
		- \Phi_{M_{5,1}^{(2)}}
	\label{eq:M51-renormalized}
\end{equation}
by integrating the convergent parametric integral $I_{M_{5,1}} - I_{M_{5,1}^{(1)}} + I_{M_{5,1}^{(2)}}$ and subtracting
\begin{equation}
	\Phi_{M_{5,1}^{(1)}}
	=0
	\quad\text{and}\quad
	\Phi_{M_{5,1}^{(2)}}
	= \Phi_{M_{3,6}} \left( \ep_7,\ep_9,\ep_4,\ep_5, \ep_6, \ep_8, \ep_3, \ep_{10} \right).
	\label{eq:M51-ct}
\end{equation}
We calculated this expansion including $\varepsilon^2$ contributions, yielding for example
\begin{align}
	&\frac{\Phi_{M_{5,1}}(1,\ldots,1)
					\cdot 4(1+3\varepsilon)
				}{G_0^4(1-2\varepsilon)^3}
	= %
\input{Expansions/M51-0000000000}
	\quad\text{and}
	\label{eq:M51-0000000000}	\\
	&\frac{\Phi_{M_{5,1}}(1+\varepsilon,\ldots,1+\varepsilon)
		\cdot 12(1+11\varepsilon)}{G_0^4(1-2\varepsilon)^3}
	= %
\input{Expansions/M51-1111111111}.
	\label{eq:M51-1111111111}
\end{align}
The first terms of the general result for arbitrary $\epe_e$ read
\begin{align}
	&\frac{\Phi_{M_{5,1}}\cdot(1+\varepsilon[3+\epe_{345678910}])(4+\epe_{345678910})}{G_0^4 (1-2\varepsilon)^3}
	=	%
\input{Expansions/M51-full},
	\label{eq:M51-full}
\end{align}
where the polynomials $p_1,p_2 \in \Q[\epe_1,\ldots,\epe_{10}]$ are given by
\begin{align}
	p_1 &=	%
\input{Expansions/M51-p_1} \\
	p_2 &= %
\input{Expansions/M51-p_2}.
	\label{eq:M51-polynomials}
\end{align}

\subsection{Symmetries}
\label{sec:4loop-symmetries}

As explained in section~\ref{sec:3loop-symmetries} we checked that our results for $\Phi_G$ are invariant under the action of $\Aut\left( G \right)$. Here we like to comment on the Fourier identity which interrelates three- and four-loop propagators.

\begin{figure}
	\begin{equation*}
		{_5 P_1} = \CloseProp{M_{3,6}} = \Graph{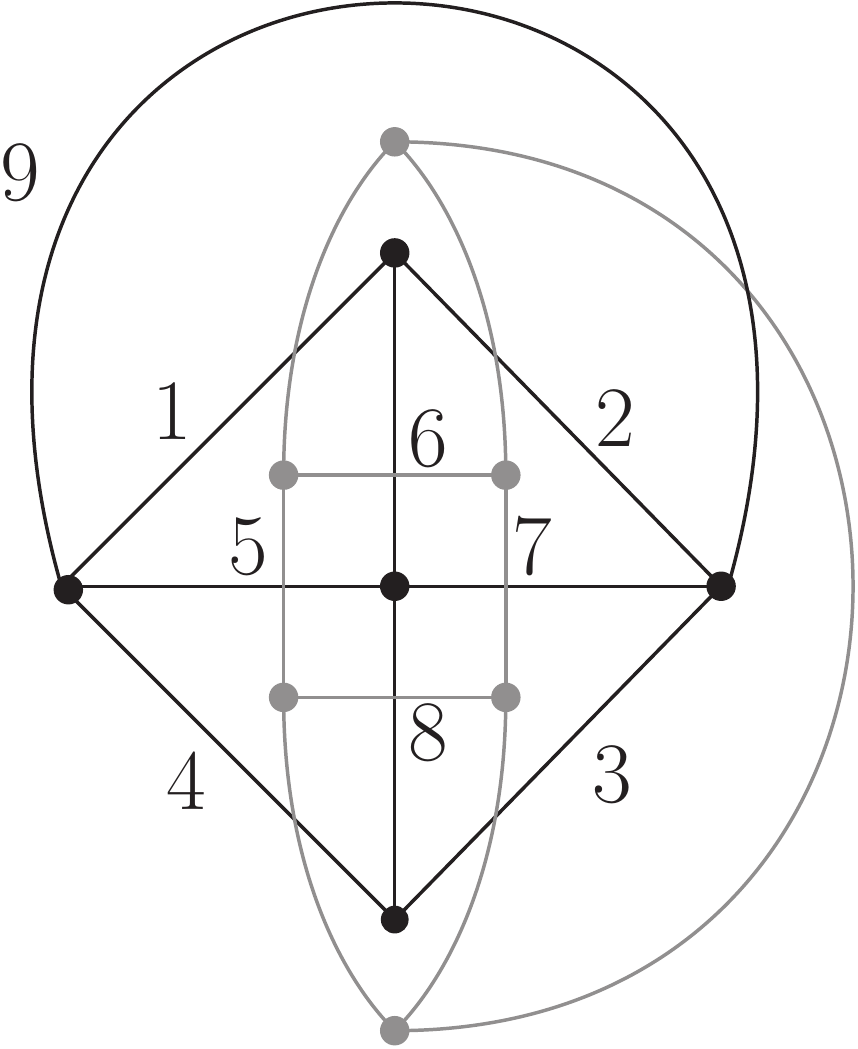}
		\quad\text{has planar dual}\quad
		Y_3 = \CloseProp{L} = \Graph{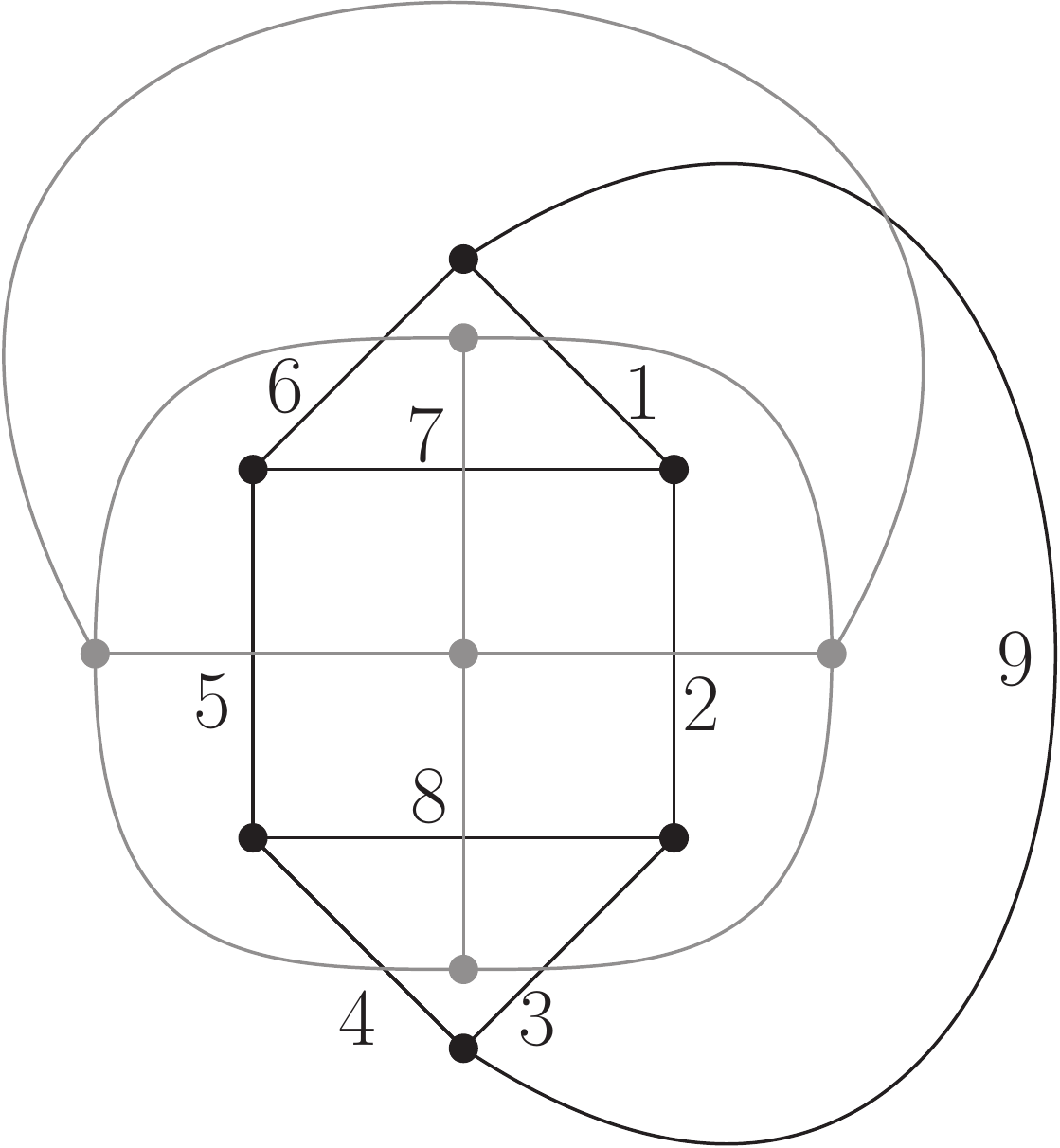}
	\end{equation*}
	\caption{%
		The planar duality $\Fourier{_5 P_1} \isomorph Y_3$ transforms the four-loop propagator $M_{3,6}$ into the three-loop propagator $L$, resulting in \eqref{eq:M36-L-duality}. Analogously we find a relation between the graphs $M_{3,5}$ and $M$ which contain subdivergences.
	}
	\label{fig:M36-L-duality}
\end{figure}

First we express the propagators in terms of their glued vacuum graphs using \eqref{eq:glueing}. We number the glued edge with nine and set $\ep_9 \defas 2- \varepsilon\left( 5 + \epe_{12345678} \right)$ in
\begin{align}
	\Phi_{M_{3,6}}
	&= \frac{\Gamma(2-\varepsilon)}{\prod_{e\in \CloseProp{E}}\Gamma(\ep_e)}
			I_{ {_5 P_1} } \left( 
				\ep_1, \ep_2, \ep_3, \ep_4, \ep_5, \ep_6, \ep_7, \ep_8, \ep_9
			\right)
	\label{eq:M36-glued}\\
	\Phi_{M_{3,5}}
	&= \frac{\Gamma(2-\varepsilon)}{\prod_{e\in \CloseProp{E}}\Gamma(\ep_e)}
			I_{ {_5 P_1} } \left( 
				\ep_1, \ep_4, \ep_7, \ep_6, \ep_2, \ep_9, \ep_3, \ep_8, \ep_5
			\right),
	\label{eq:M35-glued}
\end{align}
while for the three-loop propagators we have	$\ep_9 \defas - \varepsilon\left( 4 + \epe_{12345678} \right)$ in
\begin{align}
	\Phi_{L}
	&= \frac{\Gamma(2-\varepsilon)}{\prod_{e\in \CloseProp{E}}\Gamma(\ep_e)}
			I_{ Y_3 } \left( 
				\ep_1, \ep_2, \ep_3, \ep_4, \ep_5, \ep_6, \ep_7, \ep_8, \ep_9
			\right)
	\label{eq:L-glued}\quad\text{and}\\
	\Phi_{M}
	&= \frac{\Gamma(2-\varepsilon)}{\prod_{e\in \CloseProp{E}}\Gamma(\ep_e)}
			I_{ Y_3 } \left( 
				\ep_8, \ep_3, \ep_4, \ep_5, \ep_1, \ep_7, \ep_2, \ep_9, \ep_6
			\right).
	\label{eq:M-glued}
\end{align}
Here we fixed a labelling $1,\ldots,9$ of the edges of ${_5 P_1}$ and $Y_3$ as shown in figure~\ref{fig:M36-L-duality}. Observe how in contrast to $V$ and $Q$ of section~\ref{sec:3loop-symmetries} we can not transform the expansions of $M$ and $L$ (or $M_{3,5}$ and $M_{3,6}$) into each other because they differ on the level of $Y_3$ by the location of the glued edge $9$ that is distinguished through its power $\ep_9 = \varepsilon\epe_9$ being expanded around zero instead of one.

By Fourier transformation, the planar duality shown in figure~\ref{fig:M36-L-duality} imposes the relation
\begin{equation}
	I_{ {_5 P_1}} (\ep_1,\ldots,\ep_9)
	= I_{Y_3} \left( 
		\Fourier{\ep_2}, \Fourier{\ep_7}, \Fourier{\ep_3}, \Fourier{\ep_4}, \Fourier{\ep_5}, \Fourier{\ep_1}, \Fourier{\ep_6}, \Fourier{\ep_8}, \Fourier{\ep_9}
	\right),
	\label{eq:M36-L-duality}
\end{equation}
where both functions depend on $\ep_1,\ldots,\ep_9$ only. So explicitly we transform
\begin{equation}
	\Fourier{\ep_e}
	= \frac{\D}{2} - \ep_e
	= 2 - \varepsilon - \ep_e
	= \frac{1}{5} \sum_{i=1}^9 \ep_i - \ep_e.
	\label{eq:M36-fourier}
\end{equation}
In particular, an edge power $\ep_e = 1 + \varepsilon \epe_e$ maps to $\Fourier{\ep_e} = 1 - \varepsilon ( 1 + \epe_e )$ and is still expanded near one. The glued edge with $\ep_e = 2 + \varepsilon \epe_e$ transforms into $\Fourier{\ep_e} = -\varepsilon(1+\epe_e)$, being expanded around zero as is the case for \eqref{eq:L-glued} and \eqref{eq:M-glued}.

Finally this way we can transform the two pairs $M_{3,5} \leftrightarrow M$ of subdivergent and $M_{3,6} \leftrightarrow L$ of primitively divergent propagators into each other. Applying this transformation to our results for $\Phi_{M_{3,5}}$ and $\Phi_{M_{3,6}}$ we indeed recovered $\Phi_M$ and $\Phi_L$ which we computed independently.

Note that we used this relation as a further check for the correctness of our implementation. Practically it renders the integration of $M_{3,5}$ and $M_{3,6}$ redundant (given that $L$ and $M$ are known for arbitrary $\epe_e$).

\section{Examples with more loops}%
\label{sec:moreloops}
High-precision numerical results for massless vertex graphs of $\varphi^4$-theory were collected in \cite{Schnetz:Census} up to eight loops and constitute a formidable testing ground for our program.
First recall that the \emph{period}
\begin{equation}
	\period{\CloseProp{G}}
	\defas
	\int \frac{\Omega}{\psipol_{\CloseProp{G}}^2}
	\urel{\eqref{eq:FI-parametric-projective}}
	\restrict{I_{\CloseProp{G}}}{\varepsilon=0}
	\label{eq:period}
\end{equation}
of a primitive vertex graph $\CloseProp{G}$ (thus $\abs{E} = 2h$) is its contribution to the beta function. Explicitly, in dimensional regularization we identify it with the pole since by \eqref{eq:FI-parametric-affine}
\begin{equation}
	\Phi_{\CloseProp{G}}
	= \frac{\period{\CloseProp{G}}}{\varepsilon h}
		+ \bigo{\varepsilon^0}
	\quad\text{for primitive vertex graphs $\CloseProp{G}$ with $\abs{E} = 2h$}.
	\label{eq:period-vertex}
\end{equation}
If we cut one edge and let $G$ denote the resulting quadratically convergent propagator with $\sdc = 1 + \bigo{\varepsilon}$ (called \emph{broken primitive divergent} in \cite{Brown:TwoPoint}) we find correspondingly
\begin{equation}
	\Phi_G
	= \period{\CloseProp{G}}
		+ \bigo{\varepsilon^1}
	\quad\text{for primitive propagators $G$ with $\abs{E} = 2h + 1$},
	\label{eq:period-propagator}
\end{equation}
since integrating out the cut edge $e$ of $\CloseProp{G}$ yields the relation (recall $\psipol_{\CloseProp{G}} = \phipol_G + \SP_e \psipol_G$)
\begin{equation}
	\restrict{I_G}{\varepsilon=0}
	= \int \frac{\Omega_G}{\psipol_G \phipol_G}
	= \int \Omega_G \int_0^{\infty} \frac{\dd \SP_{e}}{(\phipol_G + \SP_{e} \psipol_G)^2}
	= \period{\CloseProp{G}}.
	\label{eq:period-propagator-vertex}
\end{equation}
In particular, the following results on periods of primitive vertex graphs are also the finite values $\restrict{\Phi_G}{\varepsilon=0}$ of all propagators obtained by cutting any internal edge.

\begin{figure}
	\begin{gather*}
		P_{6,2} = \Graph{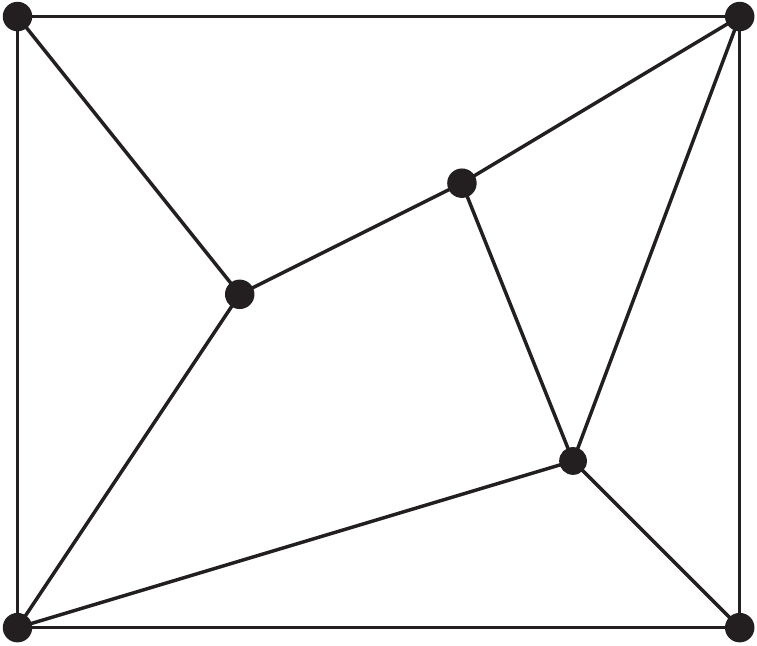}
		\quad
		P_{6,3} = \Graph{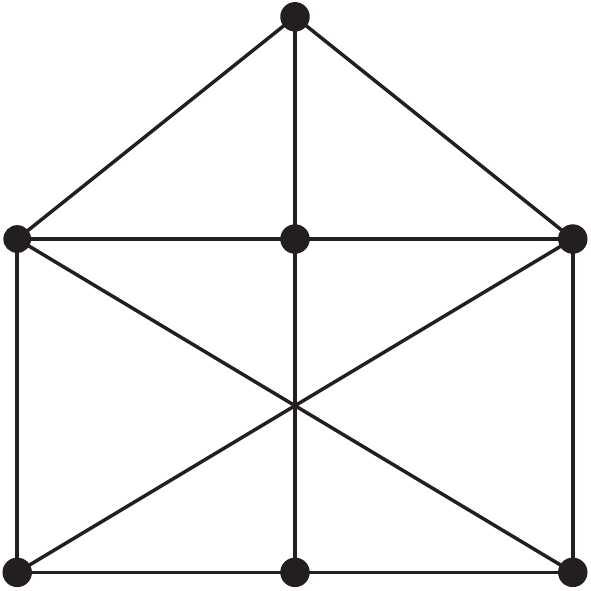}
		\quad
		P_{7,2} = \Graph{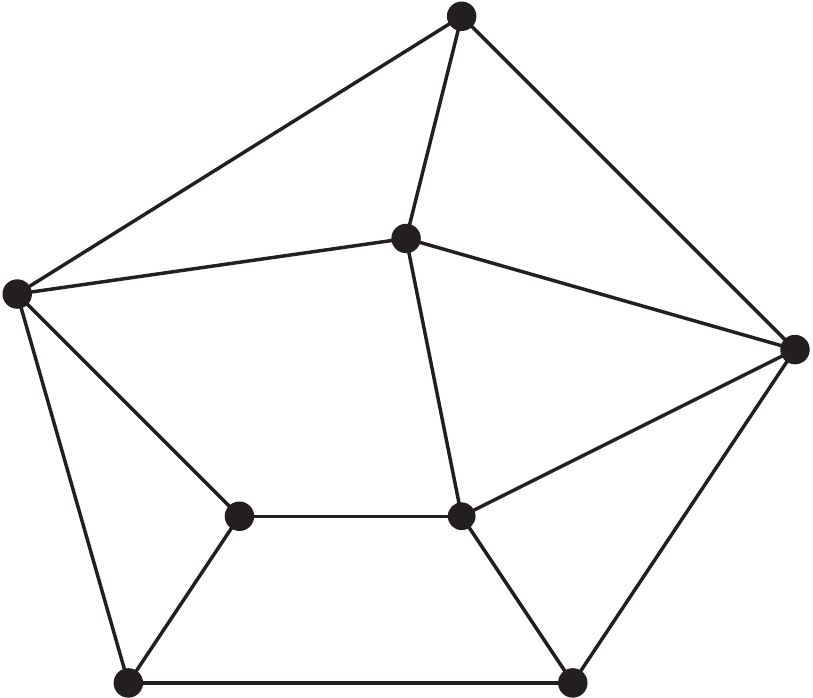}
		\quad
		P_{7,5} = \Graph{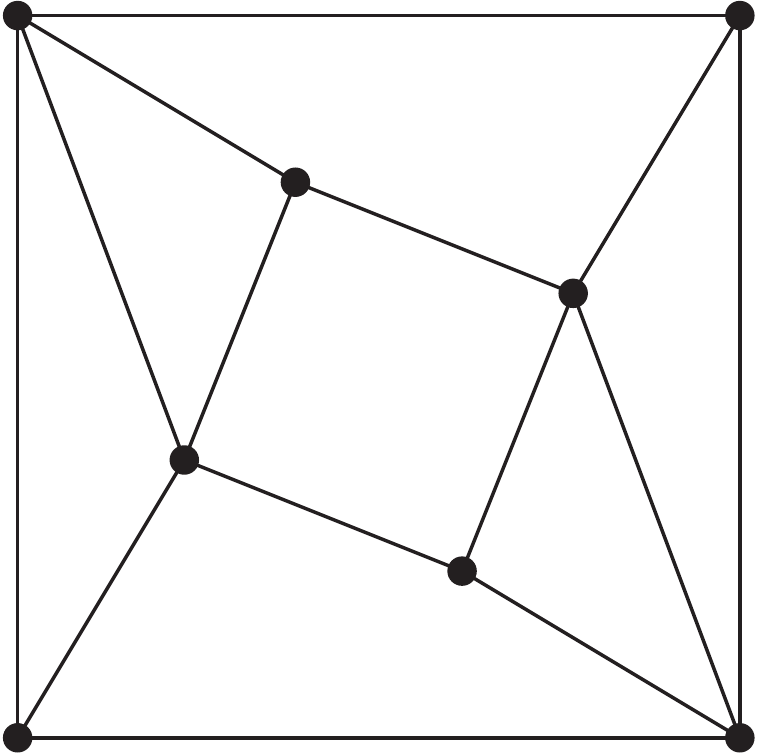}
	\end{gather*}
	\caption{%
	The six and seven loop primitive $\varphi^4$ vertex graphs from \cite{Schnetz:Census} whose periods \eqref{eq:period} we calculated in \eqref{eq:P_6,2} to \eqref{eq:P_7,5}.
	Cutting any edge of these creates five- and six-loop propagators with their leading $\varepsilon^0$-coefficient given through \eqref{eq:period-propagator}.}
	\label{fig:census}
\end{figure}

We successfully checked our implementation on the well-known wheel with $h$ spokes graphs $W_h$ and the recently evaluated \cite{BrownSchnetz:ZigZag} zig-zag graphs $Z_h$, up to $h=7$ loops where these periods are rational multiples of $\zeta_{11}^{}$.
Further we computed the graphs of figure~\ref{fig:census} and thus proved
\begin{align}
	\period{P_{6,2}}
	&= 8\zeta_3^3 + \tfrac{1063}{9} \zeta_9^{}
	\label{eq:P_6,2} \\
	\period{P_{6,3}}
	&= 252 \zeta_3^{} \zeta_5^{}
		+ \tfrac{432}{5} \zeta_{3,5}^{}
		- \tfrac{25056}{875} \zeta_2^4 
	\label{eq:P_6,3} \\
	\period{P_{7,2}}
	&= 18 \zeta_{3,3,5}^{}
		+ 35 \zeta_3^2 \zeta_5^{}
		+ 810 \zeta_2^{} \zeta_9^{}
		+ \tfrac{108}{5} \zeta_2^2 \zeta_7^{}
		- \tfrac{72}{7} \zeta_2^3 \zeta_5^{}
		- \tfrac{195379}{192} \zeta_{11}^{}
	\label{eq:P_7,2} \\
	\period{P_{7,5}}
	&= 450 \zeta_5^2 -189 \zeta_3^{} \zeta_7^{},
	\label{eq:P_7,5}
\end{align}
which confirm the numerical data in \cite{Schnetz:Census}.
Note that \eqref{eq:P_6,2} and \eqref{eq:P_7,2} are also computable\footnote{In the notation of this reference, $P_{6,2} = A_{2,0} = B_{2,0}$ and $P_{7,2} = B_{2,1} = B_{3,0} = A_{3,0}$.} with the help of \emph{graphical functions} \cite{Schnetz:GraphicalFunctions}, the value of \eqref{eq:P_6,2} has been known since \cite{BroadhurstKreimer:KnotsNumbers}.

However, to our knowledge the results \eqref{eq:P_6,3} and \eqref{eq:P_7,5} were not known analytically before. They feature a drop in transcendental weight below the typical $2h - 3$.

\subsection{Subdivergences and co-commutative graphs}
The presence of subdivergences makes renormalization necessary, which in the parametric representation is encoded in the \emph{forest formula} of \cite{BrownKreimer:AnglesScales}.
In the case of a $\varphi^4$ vertex graph $G$ with a single vertex subdivergence $\gamma$ and thus only one additional forest, the renormalized amplitude in $\D=4$ dimensions ($\varepsilon=0$) becomes
\begin{equation}
	\int \Omega \left[
		\frac{1}{\psipol_G^{2}} \ln \frac{\phipol_G}{\widetilde{\phipol}_G}
	- \frac{1}{\psipol_{\gamma}^{2} \psipol_{G/\gamma}^{2}}
			\ln \frac{
					\phipol_{G/\gamma} \psipol_{\gamma}
					+ \widetilde{\phipol}_{\gamma} \psipol_{G/\gamma}
			}{
					\widetilde{\phipol}_{G/\gamma} \psipol_{\gamma}
					+ \widetilde{\phipol}_{\gamma} \psipol_{G/\gamma}
			}
	\right]
	\label{eq:subdiv-amplitude}
\end{equation}
where $G/\gamma$ is obtained from $G$ by contracting $\gamma$ to a single vertex and the polynomial $\phipol$ now depends on masses and all external momenta.
By $\widetilde{\phipol}$ we denote its value at the chosen renormalization point. The definition \eqref{eq:period} of the period is extended to $G$ as
\begin{equation}
	\period{G} = \int \Omega \left[ 
			\frac{1}{\psipol_G^2}
			- \frac{1}{\psipol_{\gamma}^2 \psipol_{G/\gamma}^2}
			\frac{
					\widetilde{\phipol}_{G/\gamma} \psipol_{\gamma}
				}{
					\widetilde{\phipol}_{G/\gamma} \psipol_{\gamma}
					+ \widetilde{\phipol}_{\gamma} \psipol_{G/\gamma}
				}
	\right]
	\label{eq:subdiv-period}
\end{equation}
and again encodes the scaling behaviour of $G$ (thus its contribution to the beta function), see \cite{Panzer:Mellin}. So in general it depends on the renormalization point, but not so for co-commutative\footnote{with respect to the renormalization Hopf algebra of Feynman graphs} $G$ where we have $\gamma \isomorph G/\gamma$ (we recommend the remarks of \cite{Kreimer:WheelsInWheels} on this situation). 

Let $\sigma: E_G \rightarrow E_G$ be a relabelling of the edges of $G$ that restricts to such an isomorphism $\sigma: E_{\gamma} \leftrightarrow E_{G/\gamma}$, where $G/\gamma = \sigma^{-1}(\gamma)$ becomes the subdivergence of $\sigma(G)$ with quotient $\sigma(G)/\sigma^{-1}(\gamma) = \gamma$. Then by changing variables according to $\sigma$,
\begin{align}
	\period{G}
	&= \frac{\period{G} + \period{\sigma(G)}}{2}
	= \frac{1}{2} \int \Omega \Bigg[ 
			\frac{1}{\psipol_G^2}
			+\frac{1}{\psipol_{\sigma(G)}^2}
			- \frac{1}{\psipol_{\gamma}^2 \psipol_{G/\gamma}^2}
			\underbrace{\frac{
					\widetilde{\phipol}_{G/\gamma} \psipol_{\gamma}
					+\widetilde{\phipol}_{\gamma} \psipol_{G/\gamma}
				}{
					\widetilde{\phipol}_{G/\gamma} \psipol_{\gamma}
					+ \widetilde{\phipol}_{\gamma} \psipol_{G/\gamma}
					}
				}_{=1}
		\Bigg]
\hide{
	\nonumber\\
	&= \frac{1}{2} \int \Omega \left[ 
			\frac{1}{\psipol_G^2}
			+\frac{1}{\psipol_{G_{\sigma}}^2}
			- \frac{1}{\psipol_{\gamma}^2 \psipol_{G/\gamma}^2}
		\right]
}
	\label{eq:period-cocommutative}
\end{align}
reveals itself as independent of the choice of renormalization point and easily applicable to our integration method. 
The first interesting example of this type in $\varphi^4$-theory appears at six loops by inserting the wheel with 3 spokes $\gamma \isomorph G/\gamma \isomorph W_3$ into itself.
Our result reads
\begin{align}\label{eq:w3-in-w3}
	\period{\Graph[0.6]{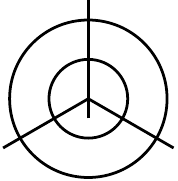}}
	&= \frac{1}{2} \int \Omega \left[
				\frac{1}{\psipol^2_{G}}
			+ \frac{1}{\psipol^2_{\sigma(G)}}
			- \frac{1}{\psipol^2_{\gamma} \psipol^2_{G/\gamma}} \right]
	= 72\zeta_3^2 - \tfrac{189}{2}\zeta_7^{}.
\end{align}
Let us now consider this graph after nullifying two of the external momenta as well as all masses to obtain a $p$-integral. 
If we renormalize by subtraction at $q^2 = \rp^2$, by \cite{Panzer:Mellin} we can read off immediately the finite amplitude at $\D=4$ from $\period{\gamma}=\period{G/\gamma}=6\zeta_3^{}$ and \eqref{eq:w3-in-w3} as
\begin{equation}
	18 \zeta_3^2 \ln^2 \frac{q^2}{\rp^2} - \left( 
		72\zeta_3^2 - \tfrac{189}{2}\zeta_7^{}
	\right) \ln \frac{q^2}{\rp^2}.
	\label{eq:w3-in-w3-renormalized}
\end{equation}
Equivalently this determines the first two terms of the un-renormalized $\varepsilon$-expansion
\begin{equation}
	\Phi_{\Graph{3Spokes}}(1,\ldots,1) \cdot G_0^{-6}
	=	2 \zeta_3^2 \varepsilon^{-2}
		+ \left(
				12 \zeta_3^2
				+ \tfrac {12}{5} \zeta_2^{2} \zeta_3^{}
				- \tfrac{63}{4} \zeta_7^{}
		\right) \varepsilon^{-1}
		+\bigo{\varepsilon^0}.
	\label{eq:w3-in-w3-dimreg}
\end{equation}
Note that \eqref{eq:period-cocommutative} generalizes whenever $G$ is cocommutative and we calculated some lower-loop examples as well. However, after fixing a renormalization point we could also integrate \eqref{eq:subdiv-period} for non-cocommutative $G$ after inserting the full forest formula. Such periods will in general depend on the choice of renormalization point though.

\section{Comments}
\label{sec:comments}

\subsection{Tensor integrals}%
\label{sec:tensors}%
In this article we restricted our considerations to scalar integrals \eqref{eq:FI-impulsraum}, but we can apply our method equally well to integrands with arbitrary products of (external- and loop-) momenta in the numerator.
We only need to transform these integrals into the parametric representation which is a simple task that can be accomplished by differentiation with respect to auxiliary variables as explained for example in \cite{KreimerSarsSuijlekom}.

The crucial property of this operation is that the parametric integrand may only acquire additional powers of the polynomial $\psipol$ in the denominator, that is, the tensor integrand in the parametric representation is $P \cdot \psipol^{-N}$ for a polynomial $P$ in the Schwinger parameters.
This implies that the polynomial reduction of section~\ref{sec:reduction} for a graph $G$ is the same no matter whether scalar or any tensor integrals are associated to it.

Hence our theorems~\ref{theorem:3loops} and \ref{theorem:4loops} immediately generalize to all subdivergence-free tensor integrals of the three- and four-loop propagators.

\subsection{Subdivergences}%
\label{sec:subdivergences}%
Hyperlogarithmic integration can only be applied to convergent integrals.
If a graph $G$ is free of subdivergences (primitive), \eqref{eq:FI-parametric-projective} is finite indeed such that we can evaluate $\Phi_G$ straightforwardly. Note that $\Gamma(\sdc)$ in \eqref{eq:FI-parametric-affine} will explicitly capture a possible overall divergence.

To evaluate $\Phi_G$ in the presence of subdivergences we need to rewrite the integral using suitable counterterms as we demonstrated in \eqref{eq:M-renormalized}, \eqref{eq:M35-renormalized} and \eqref{eq:M51-renormalized}.
This procedure effectively expresses $\Phi_G$ as a sum of products of primitive combinations of graphs which we can compute. For the example $G=M$ given in \eqref{eq:M-renormalized} and \eqref{eq:Mct2} it reads
\begin{equation}
	\Phi_M
	= \Big[
			\Phi_M - \Phi_{M^{(1)}} + \Phi_{M^{(2)}}
		\Big]
		- \Phi_{\Graph{1}} \Big(
				\ep_3,
				1+\varepsilon(2+\epe_{12678})
			\Big)
			\cdot
			\Phi_F(\ep_1,\ep_2,\ep_8,\ep_6,\ep_7),
	\label{eq:M-renormalized2}
\end{equation}
with the term in square brackets being free of subdivergences and thus amenable to our integration method.

We therefore consider our calculations of $p$-integrals rather as an abuse of the algorithm, since it is ideally suited to compute finite renormalized quantities (like amplitudes or beta functions) directly in the first place.
Note the contrast to the artificial detour taken by the common practice to first express those convergent integrals in terms of typically divergent $p$-integrals, which then need to be renormalized themselves anyway in order to be computed.

\paragraph{Weight-drops of pole coefficients}
In view of the above decomposition like \eqref{eq:M-renormalized2} we can understand drops in transcendental weight among coefficients of poles $\varepsilon^{-n}$ as observed in \cite{BaikovChetyrkin:FourLoopPropagatorsAlgebraic}:
Such terms can only arise from an at least $n$-fold product $\prod_{\gamma} \Phi_{\gamma}$ of primitive graphs $\gamma$ since each contributes at most a single pole through $\Gamma(\sdc_{\gamma})$.
The weight of $\restrict{I_{\gamma}}{\varepsilon=0}$ is at most $E_{\gamma} - 2$ since (recall that one edge $e$ is fixed to $\SP_e = 1$ and not integrated over)
\begin{enumerate}
	\item for $E_{\gamma} \in \set{ 2 h_{\gamma}, 2 h_{\gamma} + 2}$ the first integration of $\psipol^{-2}$ or respectively $\phipol^{-2}$ delivers a rational function to be integrated over a further $E_{\gamma}-2$ variables

	\item	when $E_{\gamma} = 2 h_{\gamma}+1$, the first integration of $(\psipol\phipol)^{-1}$ yields a logarithm divided by the square $W^2$ of a \emph{Dodson polynomial}\footnote{%
								We recommend to read \cite{Brown:PeriodsFeynmanIntegrals,BrownYeats:SpanningForestPolynomials} for more information on weight-drops and Dodgson polynomials.
		} $W$. Now a partial integration in the next step leaves us with an integrand still of weight one and $E_{\gamma}-3$ left over integrations.
\end{enumerate}
Therefore if we increase the order $n$ of the $\varepsilon$-pole considered, this upper bound on the transcendental weight of its coefficient decreases by two.

\paragraph{Periods of subdivergent graphs}
When $G$ contains subdivergences, the polynomial reduction of section~\ref{sec:reduction} has to be applied not only to $G$ and the counterterms individually. We rather have to calculate the singularities for all these pieces (that combine into a convergent parametric integral) of the integrand jointly.

This comes about as in \eqref{eq:cg-recursion} we intersect the singularities of the partial integrals over all possible orders of integration, which is only valid if the integral under consideration is actually convergent. 
For the example $G=M_{3,5}$ we should thus replace \eqref{eq:initial-singularities} with
\begin{equation}
	S_{\emptyset}
	\defas \set{\psipol_{M_{3,5}}, \phipol_{M_{3,5}}, \psipol_{M_{3,5}^{(1)}}, \phipol_{M_{3,5}^{(1)}}}
	\quad
	C_{\emptyset}
	\defas \set{
			\set{\psipol_{M_{3,5}}, \phipol_{M_{3,5}}},
			\set{\psipol_{M_{3,5}^{(1)}}, \phipol_{M_{3,5}^{(1)}}}
		}.
	\label{eq:M35-joint-reduction}
\end{equation}
In the very simple cases \eqref{eq:M-renormalized}, \eqref{eq:M35-renormalized} and \eqref{eq:M51-renormalized} we considered this turns out to not produce more singularities than the individual reductions.

However it is not clear to us how this will affect the linear reducibility and periodic content of graphs with subdivergences in general.
The geometry in this situation thus appears less understood and makes further study necessary.

\subsection{Constraining graph periods}
A great power of integration in parametric space lies in its predictiveness on the class of periods that will occur, to arbitrary order in the $\varepsilon$-expansion.
We are not aware of a different method that can prove results as general as theorems~\ref{theorem:3loops} and \ref{theorem:4loops} (only for the two-loop graph $F$ it was obtained in \cite{BierenbaumWeinzierl:TwoPoint} using Mellin-Barnes techniques).

Still it seems that we might further improve on these bounds, considering the striking observation that for any of the graphs $\set{N, M_{4,5},M_{5,1}}$ all coefficients we computed reduce to {\MZV}, despite the fact that our hyperlogarithmic integration algorithm produces alternating Euler sums as output.

This motivates a deeper study of the geometry of these graphs and and we very much hope to return to this question in the future.

\subsection{Efficiency}%
\label{sec:efficiency}
Apart from the theoretical results, we hope to have made clear the practical power and utility of hyperlogarithmic integration. All of our results were obtained with an implementation of \cite{Brown:TwoPoint} in {\Maple}, version 16 and computations ran single-threadedly on a 2.6 GHz machine. Some timings are reported in tables~\ref{tab:2loop-performance} and \ref{tab:N-performance}.

We stress that our foremost aim when programming was correctness and we commented on the plethora of checks we carried out along the lines of this article. 
Note that once the polynomial reduction of a graph $G$ is available, the algorithm breaks down to partial fractioning, Taylor expansions of rational functions and symbolic manipulation of words (with letters indexed by polynomials). 
In particular no numerical approximations are needed at any stage.

Therefore the hyperlogarithmic integration is completely combinatorial and we expect substantial gain in speed and reduction of memory requirements to be possible through a careful implementation in a low-level programming language, aiming for efficiency.

Having now verified its correctness and practicability, we plan to comment on and publish our current program in a separate work.

\subsection{Comparison with other methods}%
\label{sec:comparison}
\paragraph{Considering graphs individually}
	We directly integrate single graphs $G$. This feature distinguishes our approach from numerical techniques like dimensional recurrence relations used in \cite{LeeSmirnov:FourLoopPropagatorsWeightTwelve}, but also the algebraic method of \cite{BaikovChetyrkin:FourLoopPropagatorsAlgebraic} both of which require an {\IBP} reduction to master integrals that is in itself a problem of considerable complexity and restricts the reachable loop order.

	Our method therefore appears to be simpler and more generally applicable, the indispensable requirement being linear reducibility.

	However, the knowledge of identities like {\IBP} and {\GaC} does not only provide consistency checks but might also be used to express graphs intractable by our method in terms of simpler ones.
	Particularly in the light of \cite{BaikovChetyrkin:FourLoopPropagatorsAlgebraic} we see that calculation of a few new coefficients can imply results for the expansions of many other graphs.
		
		So combining these known identities with our algorithm might yield the most efficient way to perform calculations.

\paragraph{Generality}
	Whenever a graph is linearly reducible we can compute the full $\varepsilon$-expansion for arbitrary powers of the propagators and arbitrary scalar products of momenta in the numerator.
	
	Currently employed methods seem to lack this power as graphs with one-scale subdivergences (like $M_{4,2}$ of figure~\ref{fig:insertions}) are considered as non-trivial master integrals. It appears that so far only insertions of $\Graph{1}$ and $F$ into each other are considered simple in the sense that they factorize like in \eqref{eq:insertions-F'}.
	
	In the view of theorems~\ref{theorem:3loops} and \ref{theorem:4loops} we can also calculate arbitrary one-scale insertions into three- and four-loop propagators. Particularly we hope that our results of section~\ref{sec:3loops} reduce computational effort in future computations.
		
\paragraph{Subdivergences}
	As was impressively demonstrated in \cite{SmirnovTentyukov:FourLoopPropagatorsNumeric,LeeSmirnov:FourLoopPropagatorsWeightTwelve} in the case of propagators, techniques like sector-decomposition and dimensional recurrence-relations are able to provide numeric approximations of very high-precision also when subdivergences are present in a graph.

	In contrast, a separate study of the subdivergences is necessary for the application of our algorithm and it is not yet clear to which extent a high number of subdivergences might effect the linear reducibility criterion and the kind of periods that can appear.

\phantomsection
\pdfbookmark[1]{References}{final-bibliography}
\bibliographystyle{elsarticle-num}
\bibliography{../qft}

\end{document}

%% file: preamble.tex
\usepackage[english]{babel}

\usepackage{amsmath,amssymb,amsthm,amsfonts}
\usepackage{exscale}

\usepackage{lmodern}

\usepackage{hyperref}

\usepackage{graphicx}

\usepackage{ifthen}

\usepackage{shuffle}

\usepackage{booktabs}

\numberwithin{equation}{section}

\newcommand{\Q}{
\mathbb{Q}
}
\newcommand{\N}{
\mathbb{N}
}
\newcommand{\Z}{
\mathbb{Z}
}
\newcommand{\R}{
\mathbb{R}
}

\newlength{\wurelwidth}
\newcommand{\urel}[2][=]{\mathrel{\mathop{#1}\limits_{\!\scalebox{0.5}{#2}\!}}}
\newcommand{\wurel}[2][=]{\mathrel{\mathop{#1}_{\!\scalebox{0.5}{\makebox[\the\wurelwidth]{#2}}\!}}}

\newcommand{\sdc}{\omega}
\newcommand{\Graph}[2][0.3]{%
\vcenter{\hbox{\includegraphics[scale=#1]{Graphs/#2}}}%
}

\newcommand{\phipol}{\varphi}
\newcommand{\psipol}{\psi}

\newcommand{\rp}{\mu} 
\newcommand{\period}[1]{\mathcal{P}\left(#1\right)}
\newcommand{\bigo}[1]{\mathcal{O}\left(#1\right)}

\newcommand{\hide}[1]{}

\newcommand{\dd}[1][]{\mathrm{d}^{#1}}

\newcommand{\restrict}[2]{%
{\left. #1 \right|}_{#2}%
}

\DeclareMathOperator{\Aut}{Aut}

\newcommand{\isomorph}{
\cong
}
\newcommand{\defas}{
\mathrel{\mathop:}=
}

\newcommand{\set}[1]{
\left\{ #1 \right\}
}
\newcommand{\setexp}[2]{
\left\{ #1\!:\ #2 \right\}
}
\newcommand{\abs}[1]{
\left\lvert #1 \right\rvert
}

\newcommand{\MZV}{\ensuremath{\mathrm{MZV}}}

\newcommand{\qft}{%
quantum field theory%
}

\newcommand{\GaC}{GaC}
\newcommand{\IBP}{IBP}
\newcommand{\Maple}{{Maple\texttrademark}}

\newcommand{\resultant}[3]{%
{\left[#1,#2\right]}_{#3}%
}
\newcommand{\discriminant}[2]{%
D_{#2}\left( #1 \right)%
}
\newcommand{\CloseProp}[1]{%
\widetilde{#1}
}
\newcommand{\Fourier}[1]{%
\overline{#1}
}
\newcommand{\ep}{a}
\newcommand{\epe}{\nu}
\newcommand{\SP}{\alpha}
\newcommand{\D}{D}

\newtheorem{theorem}{Theorem}[section]
\newtheorem{definition}[theorem]{Definition}

\newtheorem{example}[theorem]{Example}
\newtheorem{remark}[theorem]{Remark}

%% file: Expansions/N-00000000.tex
20 \zeta_{5}^{}
+\left(\tfrac{80}{7} \zeta_{2}^{3} + 68 \zeta_{3}^{2}\right)\varepsilon
+\left(\tfrac{408}{5} \zeta_{3}^{} \zeta_{2}^{2} + 450 \zeta_{7}^{}\right)\varepsilon^{2} 
+\left(\tfrac{102228}{125} \zeta_{2}^{4} \right. \nonumber\\
\left. -2448 \zeta_{3}^{} \zeta_{5}^{} -\tfrac{9072}{5} \zeta_{3,5}^{}\right)\varepsilon^{3}
+\left(\tfrac{88036}{9} \zeta_{9}^{} -\tfrac{4640}{3} \zeta_{3}^{3} -\tfrac{10336}{7} \zeta_{2}^{3} \zeta_{3}^{} +\tfrac{19872}{5} \zeta_{2}^{2} \zeta_{5}^{}\right)\varepsilon^{4}
+\bigo{\varepsilon^{5}}

%% file: Expansions/V-0000000.tex
 20 \zeta_{5}^{}
+\left(\tfrac{80}{7} \zeta_{2}^{3} -4 \zeta_{3}^{2}\right)\varepsilon
+\left( -\tfrac{24}{5} \zeta_{2}^{2} \zeta_{3}^{} + 359 \zeta_{7}^{}\right)\varepsilon^{2}
\\&\quad 
+\left(\tfrac{90936}{875} \zeta_{2}^{4} +\tfrac{648}{5} \zeta_{3,5}^{} -56 \zeta_{3}^{} \zeta_{5}^{}\right)\varepsilon^{3}
+\left(\tfrac{248}{3} \zeta_{3}^{3} -\tfrac{192}{7} \zeta_{2}^{3} \zeta_{3}^{} -\tfrac{2112}{5} \zeta_{2}^{2} \zeta_{5}^{} +\tfrac{15364}{3} \zeta_{9}^{}\right)\varepsilon^{4}
+\bigo{\varepsilon^{5}}

%% file: Expansions/Q-0000000.tex
 20 \zeta_{5}^{}
+\left(\tfrac{80}{7} \zeta_{2}^{3} + 20 \zeta_{3}^{2}\right)\varepsilon
+\left( 24 \zeta_{2}^{2} \zeta_{3}^{} + 380 \zeta_{7}^{}\right)\varepsilon^{2}
\\&\quad
+\left(\tfrac{828}{5} \zeta_{2}^{4} + 160 \zeta_{3}^{} \zeta_{5}^{}\right)\varepsilon^{3}
+\left(\tfrac{17500}{3} \zeta_{9}^{} +\tfrac{480}{7} \zeta_{2}^{3} \zeta_{3}^{} + 96 \zeta_{2}^{2} \zeta_{5}^{} -\tfrac{880}{3} \zeta_{3}^{3}\right)\varepsilon^{4}
+\bigo{\varepsilon^{5}}

%% file: Expansions/V-1111111.tex
 20 \zeta_{5}^{}
+\left( -28 \zeta_{3}^{2} +\tfrac{80}{7} \zeta_{2}^{3}\right)\varepsilon
+\left( -\tfrac{168}{5} \zeta_{2}^{2} \zeta_{3}^{} + 2830 \zeta_{7}^{}\right)\varepsilon^{2}
\\&\quad
+\left(\tfrac{933276}{875} \zeta_{2}^{4} -10944 \zeta_{3}^{} \zeta_{5}^{} +\tfrac{3888}{5} \zeta_{3,5}^{}\right)\varepsilon^{3}
+\left( 9648 \zeta_{3}^{3} -\tfrac{43552}{7} \zeta_{2}^{3} \zeta_{3}^{} -\tfrac{44496}{5} \zeta_{2}^{2} \zeta_{5}^{} +\tfrac{921676}{3} \zeta_{9}^{}\right)\varepsilon^{4}
+\bigo{\varepsilon^{5}}

%% file: Expansions/Q-1111111.tex
 20 \zeta_{5}^{}
+\left(\tfrac{80}{7} \zeta_{2}^{3} + 44 \zeta_{3}^{2}\right)\varepsilon
+\left( 2893 \zeta_{7}^{} +\tfrac{264}{5} \zeta_{2}^{2} \zeta_{3}^{}\right)\varepsilon^{2}
\\&\quad
+\left(\tfrac{1095168}{875} \zeta_{2}^{4} +\tfrac{1944}{5} \zeta_{3,5}^{} -120 \zeta_{3}^{} \zeta_{5}^{}\right)\varepsilon^{3}
+\left(\tfrac{956684}{3} \zeta_{9}^{} -\tfrac{832}{7} \zeta_{2}^{3} \zeta_{3}^{} -\tfrac{6192}{5} \zeta_{2}^{2} \zeta_{5}^{} -20376 \zeta_{3}^{3}\right)\varepsilon^{4}
+\bigo{\varepsilon^{5}}

%% file: Expansions/Q-full.tex
20 \zeta_5^{}
+ \left\{
	\tfrac{80}{7} \zeta_2^3
	+2 \zeta_3^2 p_1
\right\}\varepsilon
+ \left\{
	\zeta_7^{} \left( 380 + 7 p_2 \right)
	+\tfrac{12}{5} \zeta_2^2 \zeta_3^{} p_1
\right\}\varepsilon^{2}
\nonumber\\&\quad
+ \left\{
	\tfrac{9}{5}\zeta_{3,5}^{} \left( p_1-10 \right)  \left( p_1-4 \right) 
	+ \zeta_2^4 \left( \cdots \right)
	+ \zeta_3^{} \zeta_5^{} \left( \cdots \right)
\right\}{\varepsilon}^{3}
+\bigo{\varepsilon^4}

%% file: Expansions/Q-p_1.tex
10+3\,\epe_{2367}

%% file: Expansions/Q-p_2.tex
16 \epe_{67}
+ \tfrac{17}{8} \epe_{2367}^{2}
+ \tfrac{69}{4} \epe_{2367}
+ 24\,\epe_{14}
+ 32\,\epe_{5}
\nonumber\\&\quad
+ 6 \left(
			\epe_{1}\epe_{127}
		+ \epe_{4}\epe_{346}
		+ \epe_{67}^{2}
		- \epe_{6}\epe_{7}
		- \epe_{2}\epe_{3}
	\right)
+ 4 \epe_{5} \left( \epe_{23}+2\,\epe_{145}+3\,\epe_{67} \right)
\nonumber\\&\quad
+ 8 \left(
			\epe_{1}\epe_{6}
		+ \epe_{4}\epe_{7}
	\right)
+ 2 \left(
			\epe_{12}\epe_{46}
		+ \epe_{17}\epe_{34}
		+ \epe_{23}^{2}
	\right)

%% file: Expansions/V-full.tex
20 \zeta_5^{}
+ \left\{
	\tfrac{80}{7} \zeta_2^3
	-2 \zeta_3^2 p_1
\right\}\varepsilon
+ \left\{
	\zeta_7^{} \left( 359+7 p_2 \right)
	-\tfrac{12}{5} \zeta_2^2 \zeta_3^{} p_1
\right\}{\varepsilon}^{2}
\nonumber\\&\quad
+ \left\{
 	\tfrac{9}{5}\zeta_{3,5}^{} \left( 4+p_{{1}} \right)  \left( 10+p_{{1}} \right)
	+\zeta_2^4 (\cdots)
	+\zeta_3^{}\zeta_5^{} \left( \cdots \right)
\right\}{\varepsilon}^{3}
+\bigo{\varepsilon^4}

%% file: Expansions/V-p_1.tex
2+3\,\epe_{4567}

%% file: Expansions/V-p_2.tex
\tfrac{1}{8}\, \epe_{4567}^{2}
-\tfrac{1}{4}\epe_{4567}
+32\,\epe_{267}
+16\,\epe_{45}
+24\,\epe_{13}
+6\,\epe_{3}\epe_{3467}+6\,\epe_{1}\epe_{1567}
\nonumber\\&\quad
+4\,\epe_{2} \left( \epe_{45}+2\,\epe_{123}+3\,\epe_{67} \right) +4\,\epe_{1}\epe_{3}
+4 \left(
			\epe_{1}\epe_{6}
		+ \epe_{3}\epe_{7}
		+ \epe_{67}^{2}
	\right)
\nonumber\\&\quad
+2 \left(
		\epe_{1}\epe_{4}
		+\epe_{3}\epe_{5}
		+\epe_{47}^{2}
		+\epe_{56}^{2}
		+\epe_{46}^{2}
		+\epe_{57}^{2}
		+\epe_{46}\epe_{57}
	\right)

%% file: Expansions/L_00000000.tex
20 \zeta_{5}^{}
+\left(
	\tfrac{80}{7} \zeta_{2}^{3}
	+ 44 \zeta_{3}^{2}
\right)\varepsilon
+\left(
		\tfrac{264}{5} \zeta_{2}^{2} \zeta_{3}^{}
		+ 317 \zeta_{7}^{}
\right)\varepsilon^{2} 
+\left( 
		1336 \zeta_{3}^{} \zeta_{5}^{}
		+\tfrac{1944}{5} \zeta_{3,5}^{}
\right.\nonumber\\&\quad\left.
		+\tfrac{13248}{875} \zeta_{2}^{4}
\right)\varepsilon^{3}
+\left(
		\tfrac{4992}{7} \zeta_{2}^{3} \zeta_{3}^{}
		-\tfrac{1824}{5} \zeta_{2}^{2} \zeta_{5}^{}
		+\tfrac{11092}{3} \zeta_{9}^{}
		-\tfrac{664}{3} \zeta_{3}^{3}
\right)\varepsilon^{4}
+\bigo{\varepsilon^{5}}

%% file: Expansions/L_11111111.tex
 20 \zeta_{5}^{}
+\left(\tfrac{80}{7} \zeta_{2}^{3} + 116 \zeta_{3}^{2}\right)\varepsilon
+\left(\tfrac{696}{5} \zeta_{2}^{2} \zeta_{3}^{} + 2200 \zeta_{7}^{}\right)\varepsilon^{2}
+\left( 19344 \zeta_{3}^{} \zeta_{5}^{} 
+\tfrac{23328}{5} \zeta_{3,5}^{} 
\right.\nonumber\\&\quad\left.
-\tfrac{413484}{875} \zeta_{2}^{4}\right)\varepsilon^{3}
+\left(\tfrac{76448}{7} \zeta_{2}^{3} \zeta_{3}^{} -\tfrac{11952}{5} \zeta_{2}^{2} \zeta_{5}^{} + 179872 \zeta_{9}^{} -26208 \zeta_{3}^{3}\right)\varepsilon^{4}
+\bigo{\varepsilon^{5}}

%% file: Expansions/L-full.tex
20 \zeta_5^{}
+ \left\{
	\tfrac{80}{7} \zeta_2^3
	+2 \zeta_3^2 p_1
\right\}\varepsilon
+ \left\{
		\tfrac{12}{5} \zeta_2^2 \zeta_3^{} p_1
		+\zeta_7 \left( 317+14 p_{{2}} \right)
\right\}\varepsilon^2
\\&\quad
+ \left\{
	\tfrac{9}{5} \zeta_{3,5}^{} \left( p_{{1}}-4 \right)  \left( p_{{1}}-10 \right) 
	+\tfrac{6}{875} \zeta_{2}^{4} \left( 
			15210 + 1323\,p_{{1}}-87\,p_{1}^{2} + 980\,p_{{2}} \right)
	+ \zeta_{3}^{} \zeta_{5}^{} \cdot p_{{3}}
\right\}\varepsilon^3
\nonumber\\&\quad
+ \left\{
		\tfrac{4}{7} \zeta_{2}^{3} \zeta_{3}^{} \left( p_{{3}}-4 p_{1} \right)
		+ \tfrac{3}{5} \zeta_{2}^{2} \zeta_{5}^{} \Big( p_{{3}}-9 ( p_1-4 )  ( p_1-10 )  \Big)
		+\zeta_{3}^{3} \cdot p_4
		+\zeta_{9}^{} \cdot p_5
\right\}\varepsilon^4
+\bigo{\varepsilon^5}

%% file: Expansions/L-p_1.tex
22+3\,\epe_{1346}+6\,\epe_{2578}

%% file: Expansions/L-p_2.tex
\epe_{12}^{2}
+ \epe_{23}^{2}
+ \epe_{45}^{2}
+ \epe_{56}^{2}
+ \epe_{17}^{2}
+ \epe_{38}^{2}
+ \epe_{48}^{2}
+ \epe_{67}^{2}
+2 \epe_{25}\epe_{78}
\nonumber\\&\quad
+\epe_{14}\epe_{36}
+\epe_{78}\epe_{134678}
-2 \epe_{78}
-6 \epe_{25}
+ \tfrac{1}{144} \left( p_{1}-22 \right) \left( p_{1} + 320 \right) 

%% file: Expansions/M-00000000.tex
- 6 \zeta_{3}^{} \varepsilon^{-1}
 -\tfrac{18}{5} \zeta_{2}^{2}
 + 138 \zeta_{5}^{}\varepsilon
+\left(\tfrac{576}{7} \zeta_{2}^{3} + 510 \zeta_{3}^{2}\right)\varepsilon^{2}
\nonumber\\&
+\left( 612 \zeta_{2}^{2} \zeta_{3}^{} + 3315 \zeta_{7}^{}\right)\varepsilon^{3}
+\left( 5460 \zeta_{3}^{} \zeta_{5}^{} +\tfrac{1738746}{875} \zeta_{2}^{4} -\tfrac{5832}{5} \zeta_{3,5}^{}\right)\varepsilon^{4}
+\bigo{\varepsilon^{5}}

%% file: Expansions/M-11111111.tex
 -6 \zeta_{3}^{}\varepsilon^{-1}
 -\tfrac{18}{5} \zeta_{2}^{2}
 + 1248 \zeta_{5}^{}\varepsilon
+\left(\tfrac{5016}{7} \zeta_{2}^{3} + 14352 \zeta_{3}^{2}\right)\varepsilon^{2}
\nonumber\\&
+\left(\tfrac{86112}{5} \zeta_{2}^{2} \zeta_{3}^{} + 239697 \zeta_{7}^{}\right)\varepsilon^{3}
+\left( 1528068 \zeta_{3}^{} \zeta_{5}^{} +\tfrac{129545532}{875} \zeta_{2}^{4} -\tfrac{442584}{5} \zeta_{3,5}^{}\right)\varepsilon^{4}
+\bigo{\varepsilon^{5}}

%% file: Expansions/M-full.tex
-6 \zeta_{3}^{} \varepsilon^{-1}
- \tfrac{18}{5} \zeta_2^2
+\zeta_{5}^{} \left( 138+5 p_1 \right) \varepsilon
\nonumber\\&\quad
+ \left\{
		\tfrac{4}{7} \zeta_2^3 \left(144 + 5\,p_{{1}}\right) 
		+ \zeta_{3}^{2} \cdot p_2
\right\} {\varepsilon}^{2}
+ \left\{
		\tfrac{6}{5} \zeta_2^2 \zeta_3^{} p_2
		+ \zeta_7^{} \left( \cdots \right)
\right\} \varepsilon^{3}
+\bigo{\varepsilon^4}

%% file: Expansions/M-p_1.tex
15 \epe_{1378}
+12 \epe_{2}
+9 \epe_{45}
+21 \epe_{6}
+ \epe_{1378}^{2}
+2 \epe_{18}\epe_{37}
\nonumber\\&\quad
+3 \left(
			\epe_{456} \epe_{123678}
		+ \epe_{2}\epe_{1378}
		+ \epe_{6}\epe_{12378}
		+ \epe_{3}\epe_{7}
		+ \epe_{1}\epe_{8}
	\right)

%% file: Expansions/N-full.tex
20 \zeta_5
+\varepsilon \left\{
	\tfrac{80}{7} \zeta_2^3
	+ \zeta_3^2 \left( 68+6 p_1 \right)
\right\}
\nonumber\\&\quad
+\varepsilon^2 \left\{
	\tfrac{6}{5} \zeta_2^2 \zeta_3 \left( 68+6 p_1 \right)
	+ \zeta_7 \left( 450 - 14 p_2 \right)
\right\}
+\bigo{\varepsilon^3}

%% file: Expansions/N-p_1.tex
2\,\epe_{1346}+3\,\epe_{2578}

%% file: Expansions/N-p_2.tex
2 \left(
	\epe_{2578}\epe_{1346}
	+ \epe_{3}\epe_{457}
	+ \epe_{4}\epe_{238}
	+ \epe_{6}\epe_{127}
	+ \epe_{1}\epe_{568}
\right)
+3 (\epe_{25}\epe_{78}+\epe_{16}\epe_{34})
\nonumber\\&\quad
+4 (\epe_{7}\epe_{8} + \epe_{2}\epe_{5})
+ \left( 1-p_1 \right) \epe_{12345678}
-\tfrac{1}{16} p_1 \left( p_1 + 90 \right) 

%% file: Expansions/N-11111111.tex
 20 \zeta_{5}^{}
+\left( 188 \zeta_{3}^{2} +\tfrac{80}{7} \zeta_{2}^{3}\right)\varepsilon
+\left(\tfrac{1128}{5} \zeta_{2}^{2} \zeta_{3}^{} + 3271 \zeta_{7}^{}\right)\varepsilon^{2}
\nonumber\\&\quad
+\left( -113832 \zeta_{3}^{} \zeta_{5}^{} +\tfrac{21820152}{875} \zeta_{2}^{4} -\tfrac{351864}{5} \zeta_{3,5}^{}\right)\varepsilon^{3}
\nonumber\\&\quad
+\left( -\tfrac{456832}{7} \zeta_{2}^{3} \zeta_{3}^{} +\tfrac{714096}{5} \zeta_{2}^{2} \zeta_{5}^{} +\tfrac{1233016}{3} \zeta_{9}^{} -71400 \zeta_{3}^{3}\right)\varepsilon^{4}
+\bigo{\varepsilon^{5}}

%% file: Expansions/M36-00000000.tex
 20 \zeta_{5}^{}\varepsilon^{-1}
 +\tfrac{80}{7} \zeta_{2}^{3} -28 \zeta_{3}^{2}
+\left( -\tfrac{168}{5} \zeta_{2}^{2} \zeta_{3}^{} + 254 \zeta_{7}^{}\right)\varepsilon
\nonumber\\&\qquad
+\left(\tfrac{3888}{5} \zeta_{3,5}^{} -\tfrac{148644}{875} \zeta_{2}^{4} + 88 \zeta_{3}^{} \zeta_{5}^{}\right)\varepsilon^{2}
+\bigo{\varepsilon^{3}}

%% file: Expansions/M36-11111111.tex
 20 \zeta_{5}^{}\varepsilon^{-1}
 +\tfrac{80}{7} \zeta_{2}^{3} -100 \zeta_{3}^{2}
+\left( 2011 \zeta_{7}^{} -120 \zeta_{2}^{2} \zeta_{3}^{}\right)\varepsilon
\nonumber\\&\qquad
+\left( 5832 \zeta_{3,5}^{} -\tfrac{179832}{175} \zeta_{2}^{4} -23648 \zeta_{3}^{} \zeta_{5}^{}\right)\varepsilon^{2}
+\bigo{\varepsilon^{3}}

%% file: Expansions/M36-full.tex
\tfrac{20}{\varepsilon}\zeta_5
+\tfrac{80}{7} \zeta_2^3
-\zeta_3^2 \left( 28+6\,p_{{1}} \right) 
\nonumber\\&
+\varepsilon \left\{
	-\tfrac {12}{5} \zeta_2^2 \zeta_3 \left( 14+3\,p_1 \right)
	+\zeta_7 \left( 
			254
			+ \tfrac{231}{8} p_1^2
			+ \tfrac{357}{4} p_1
			-14 p_2
	\right)
\right\}
+\varepsilon^{2} \left\{
	\tfrac{81}{5} \zeta_{3,5}
		 \left( 6+p_{{1}} \right)  \left( 8+p_{{1}} \right)
\right.\nonumber\\&\left.
 -\tfrac{3}{125} \zeta_2^4 \left(
		 	\tfrac {49548}{7}
			+1437\,p_{{1}}
			-\tfrac {4953}{14} p_1^2
			+280 p_2
	\right)
	+ \zeta_3 \zeta_5 \left(
			88
			+3 p_1^3
			+2 p_1 p_2
\right.\right.\nonumber\\&\left.\left.
			-20 \epe_{12345678} p_1^2
			+29 p_1^2
			-240 p_1 \epe_{12345678}
			+28 p_2
			+20 p_3
			-546 p_1
	 \right)
 \right\} 
 +\bigo{\varepsilon^3}

%% file: Expansions/M36-p_1.tex
\epe_{1234}+2\,\epe_{5678}

%% file: Expansions/M36-p_2.tex
6\,\epe_{5678}\epe_{12345678}+2\,\epe_{5}\epe_{23678}+2\,\epe_{7}
\epe_{14568}-\epe_{68}\epe_{123468}+2\,\epe_{6}\epe_{348}+2\,\epe_{8}
\epe_{126}
\nonumber\\&\quad
+3\,\epe_{13}\epe_{24}+4\,\epe_{2}\epe_{4}+4\,\epe_{1}
\epe_{3}+2\,\epe_{68}+6\,\epe_{57}

%% file: Expansions/M36-p_3.tex
 \left( \epe_{1234}+4 \right)  \left( \epe_{1}-\epe_{3} \right) 
 \left( \epe_{2}-\epe_{4} \right)
 +2\,\epe_{5}\epe_{7} \left( 3+\epe_{1234}+2\,\epe_{68} \right)
-2\,\epe_{6}\epe_{8}
+2\,\epe_{1234}\epe_{68}
\nonumber\\&\quad
+ \left( \epe_{5}-\epe_{7} \right)
\left( \epe_2^2+ \epe_3^2- \epe_1^2-\epe_4^2 \right) 
+ \epe_{68}^{2}\epe_{1234}
+2\,\epe_{6}\epe_{8}\epe_{57}
+ \left( 4\,\epe_{68}+6 \right) \left( \epe_{5}\epe_{23}+\epe_{7}\epe_{14} \right)
\nonumber\\&\quad
+2\,\epe_{6} \left( \epe_{3}\epe_{147}+\epe_{4}\epe_{25}+2\,\epe_{1}\epe_{2}+\epe_{12}
 \right) 
+2\,\epe_{8} \left( \epe_{2}\epe_{147}+\epe_{1}\epe_{35}+2\,
\epe_{3}\epe_{4}+\epe_{34} \right)
\nonumber\\&\quad
+2 \left(
 		\epe_{1}\epe_{4}\epe_{7}
	 +\epe_{2}\epe_{3}\epe_{5}
		+\epe_{57}\epe_{145}\epe_{237}
\right)
+36\,\epe_{5678}

%% file: Expansions/M44-full.tex
{\tfrac {441}{8}}\,\zeta_7
+\varepsilon\, \left\{
	{\tfrac {55701}{1750}}\,\zeta_2^4
	-{\tfrac {81}{5}}\zeta_{3,5}
	-\zeta_3 \zeta_5 \left( 
		135
		+ 9 p_1
	\right)
	\right\}
+{\varepsilon}^{2} \left\{
	-{\tfrac {27}{5}} \zeta_2^2\zeta_5 \left( 6+p_1 \right)
\right.\nonumber\\&\quad\left.
	-\zeta_3^3 \left( 267 + 21 p_1 + 4 p_2 \right)
	-{\tfrac{36}{7}} \zeta_2^3\zeta_3  \left( 15+p_{{1}} \right)
	+\zeta_9 \left( \tfrac{4583}{2}+\tfrac{p_3}{36}\right)
\right\}
+\bigo{\varepsilon^3}

%% file: Expansions/M44-p_1.tex
7\,\epe_{8}+2\,\epe_{36}+3\,\epe_{14}+4\,\epe_{25}+5\,\epe_{79}

%% file: Expansions/M44-p_2.tex
2\,\epe_{123789}\epe_{1278}+2\,\epe_{456789}\epe_{4589}+\epe_{14}-
\epe_{789}-2\,\epe_{2356}

%% file: Expansions/M44-p_3.tex
7435 \epe_{7}\epe_{9}
+1674 \epe_{1}\epe_{4}
+4252 (\epe_{1}\epe_{2}
			+\epe_{4}\epe_{5})
+4197 (\epe_{3}^{2}
				+\epe_{6}^{2} )
+6378 \epe_{8}^{2}
+55 ( \epe_{7}^{2}
			+\epe_{9}^{2} )
\nonumber\\&\quad
+5706 (\epe_{1}\epe_{68}
				+\epe_{38}\epe_{4})
+1509 \left( \epe_{38}+2\,\epe_{2} \right)
			\left( \epe_{68}+2\,\epe_{5} \right)
+4142 ( \epe_{1} ^{2}
				+ \epe_{4}^{2} )
\nonumber\\&\quad
+3800 (\epe_{1}\epe_{9}
			+\epe_{4}\epe_{7})
+4527 (\epe_{1}\epe_{5}
			+\epe_{2}\epe_{4})
+7270 (\epe_{5}\epe_{89}
				+\epe_{2}\epe_{78})
+220 \epe_{8}\epe_{14789}
\nonumber\\&\quad
+7105 (\epe_{167}\epe_{7}
				+\epe_{349}\epe_{9})
+3635 (\epe_{3}\epe_{78}
				+\epe_{6} \epe_{89} )
+6323 (\epe_{5}\epe_{56}
				+\epe_{2}\epe_{23})
\nonumber\\&\quad
+110 (\epe_{3}\epe_{9}
			+\epe_{6}\epe_{7})
+2126 (\epe_{1}\epe_{3}
				+\epe_{4}\epe_{6})
+10850 (\epe_{28}\epe_{9}
				+\epe_{58}\epe_{7})
\nonumber\\&\quad
+ {\tfrac {73525}{2}}\,\epe_{79}
+32935\,\epe_{25}
+{\tfrac {43895}{2}}\,\epe_{14}
+21700\,\epe_{36}
+{\tfrac {82225}{2}}\,\epe_{8}

%% file: Expansions/M44_000000000.tex
\tfrac{441}{8} \zeta_{7}^{}
+\left(\tfrac{55701}{1750} \zeta_{2}^{4} -\tfrac{81}{5} \zeta_{3,5}^{} -135 \zeta_{3}^{} \zeta_{5}^{}\right)\varepsilon 
\nonumber\\&\quad
+\left(\tfrac{4583}{2} \zeta_{9}^{} -\tfrac{540}{7} \zeta_{2}^{3} \zeta_{3}^{} -\tfrac{162}{5} \zeta_{2}^{2} \zeta_{5}^{} -267 \zeta_{3}^{3}\right)\varepsilon^{2}
+\bigo{\varepsilon^{3}}

%% file: Expansions/M44_111111111.tex
\tfrac{441}{8} \zeta_{7}^{}
+\left(\tfrac{55701}{1750} \zeta_{2}^{4} -\tfrac{81}{5} \zeta_{3,5}^{} -450 \zeta_{3}^{} \zeta_{5}^{}\right)\varepsilon
\nonumber\\&\qquad
+\left(\tfrac{410141}{24} \zeta_{9}^{} -1350 \zeta_{3}^{3} -\tfrac{1800}{7} \zeta_{2}^{3} \zeta_{3}^{} -\tfrac{1107}{5} \zeta_{2}^{2} \zeta_{5}^{}\right)\varepsilon^{2}
+\bigo{\varepsilon^{3}}

%% file: Expansions/M45-full.tex
36 \zeta_3^2
+\varepsilon \left\{
	\tfrac{216}{5} \zeta_2^2\zeta_3^{}
	-\tfrac{189}{2} \zeta_7^{} \left(4+p_1 \right)
\right\}
+\varepsilon^{2} \left\{
	\tfrac{27}{5} \zeta_{3,5}^{} \left(
		112
		+44 p_{{1}}
		+16 p_{{2}}
	\right)
\right.\nonumber\\&\left.
	-\tfrac{54}{875} \zeta_2^4 \left(
			5978
			+2011 p_{{1}}
			+464 p_{{2}}
		\right)
	+6 \zeta_3^{}\zeta_5^{} \left(
		474
		+27 p_2
		+12 p_1^2
		+5 p_3
	\right)
\right\}
+\bigo{\varepsilon^3}

%% file: Expansions/M45-p_1.tex
\epe_{234578}+2\,\epe_{9}

%% file: Expansions/M45-p_2.tex
\epe_{2389}\epe_{4579}

%% file: Expansions/M45-p_3.tex
10\,\epe_{16}+40\,\epe_{34}+45\,\epe_{2578}+74\,\epe_{9}
-\epe_{9}^{2}
+\epe_{34}\epe_{1256789}
-\epe_{2}\epe_{7}
-\epe_{3}\epe_{4}
-\epe_{5}\epe_{8}
\nonumber\\&\quad
+3 (\epe_{58}\epe_{568} + \epe_{27}\epe_{127})
+2 (\epe_{34}^{2} + \epe_{1}^{2} + \epe_{6}^{2}+ \epe_{1279} \epe_{5689})

%% file: Expansions/M45-000000000.tex
36 \zeta_{3}^{2}
+\left( 
	\tfrac{216}{5} \zeta_{2}^{2} \zeta_{3}^{}
	-378 \zeta_{7}^{}
\right)\varepsilon
\nonumber\\&\quad
+\left( 
	2844 \zeta_{3}^{} \zeta_{5}^{}
	-\tfrac{46116}{125} \zeta_{2}^{4}
	+\tfrac{3024}{5} \zeta_{3,5}^{}
\right)\varepsilon^{2}
+\bigo{\varepsilon^{3}}

%% file: Expansions/M35-00000000.tex
 6 \zeta_{3}^{}\varepsilon^{-2}
 +\tfrac{18}{5} \zeta_{2}^{2}\varepsilon^{-1}
 -138 \zeta_{5}^{}
+\left( -\tfrac{576}{7} \zeta_{2}^{3} + 174 \zeta_{3}^{2}\right)\varepsilon
\nonumber\\&
+\left(\tfrac{1044}{5} \zeta_{2}^{2} \zeta_{3}^{} -3315 \zeta_{7}^{}\right)\varepsilon^{2}
+\left( 13836 \zeta_{3}^{} \zeta_{5}^{} +\tfrac{5832}{5} \zeta_{3,5}^{} -\tfrac{1523286}{875} \zeta_{2}^{4}\right)\varepsilon^{3}
+\bigo{\varepsilon^{4}}

%% file: Expansions/M35-11111111.tex
 6 \zeta_{3}^{}\varepsilon^{-2}
 +\tfrac{18}{5} \zeta_{2}^{2}\varepsilon^{-1}
 -1248 \zeta_{5}^{}
+\left( -\tfrac{5016}{7} \zeta_{2}^{3} + 6252 \zeta_{3}^{2}\right)\varepsilon
\nonumber\\&
+\left(\tfrac{37512}{5} \zeta_{2}^{2} \zeta_{3}^{} -239697 \zeta_{7}^{}\right)\varepsilon^{2}
+\left( 2890128 \zeta_{3}^{} \zeta_{5}^{} +\tfrac{442584}{5} \zeta_{3,5}^{} -\tfrac{123055272}{875} \zeta_{2}^{4}\right)\varepsilon^{3}
+\bigo{\varepsilon^{4}}

%% file: Expansions/M35-total.tex
{\frac {6\zeta_3^{}}{{\varepsilon}^{2}}}
+\frac {18\zeta_2^2}{5\varepsilon}
-\zeta_5^{} \left( 138+15\,p_{{1}} \right)
\nonumber\\&
+\varepsilon \left\{
	-\frac{\zeta_2^3}{7} \left( 576+60\,p_{{1}} \right)
	+ \zeta_3^2 \left( 174+3\,p_{{1}}+6\,p_{{2}} \right)
\right\}
+\bigo{\varepsilon^2}

%% file: Expansions/M35-p_1.tex
\tfrac{1}{3}	\epe_{2367}^2
+\tfrac{2}{3} \epe_{27}\epe_{36}
+\epe_{2}\epe_{7}
+\epe_{3}\epe_{6}
+7 \epe_{5}
+5 \epe_{2367}
+3 \epe_{14}
+4 \epe_{8}
\nonumber\\&\quad
+ \left( 2\,\epe_{5}+\epe_{148} \right) \epe_{2367}
+\epe_{58} \epe_{1458}
- \epe_{8}^{2}

%% file: Expansions/M35-p_2.tex
8\,\epe_{238}
+32\,\epe_{67}
+3\,\epe_{2}\epe_{7}+3\,\epe_{3}\epe_{6}
+2\,\epe_{67}\epe_{2}\epe_{3}
-2\,\epe_{1}\epe_{4}\epe_{12345678}
+2\,\epe_{5}\epe_{23678} \left( \epe_{238}+3\,\epe_{67} \right)
\nonumber\\&\quad
+13\,\epe_{67}\epe_{8}
+3\, \epe_{67}^{2}\epe_{238}
+\epe_{8} \left( 2\,\epe_{6}\epe_{7}+\epe_{8} \left( \epe_{67}-\epe_{23} \right)  \right)
+\epe_{2} \left( 2\,\epe_{7}\epe_{8}- \epe_{6}^{2} \right)
+\epe_{3} \left( 2\,\epe_{6}\epe_{8}- \epe_{7}^{2} \right)
\nonumber\\&\quad
+ \epe_{5}^{2} \left( 6\,\epe_{238}+8\,\epe_{67}+4\,\epe_{5} \right) 
+2\,\epe_{6}\epe_{7}
+ \epe_{14}^{2}\epe_{235678}
-8\,\epe_{1}\epe_{4}
+\epe_{23} \left( 22\,\epe_{5}-\epe_{8}+10\,\epe_{67} \right)
\nonumber\\&\quad
+\epe_{14} \left( \epe_{567} \left( 3\,\epe_{67}+5\,\epe_{5}
 \right) +\epe_{238} \left( \epe_{238}+4\,\epe_{67}+6\,\epe_{5}
 \right)  \right) 
 +\epe_{5} \left( 26\,\epe_{5}+36\,\epe_{67}+22\,\epe_{8} \right) 
\nonumber\\&\quad
 +\epe_{14} \left( \epe_{14}+9\,\epe_{238}+15\,\epe_{67}+21\,\epe_{5} \right)
+2\,\epe_{2}\epe_{3}
+ \epe_{2}^{2} \left( \epe_{7}-\epe_{8} \right)
 +\epe_{3}^{2} \left( \epe_{6}-\epe_{8} \right)
\nonumber\\&\quad
 +2 \epe_{67}^{3}
 +14 \epe_{67}^{2}
 +14\,\epe_{14}
 +50\,\epe_{5}

%% file: Expansions/M51-0000000000.tex
 -20 \zeta_{5}^{}\varepsilon^{-1}
 -\tfrac{80}{7} \zeta_{2}^{3} -68 \zeta_{3}^{2}
+\left( -\tfrac{408}{5} \zeta_{2}^{2} \zeta_{3}^{} -170 \zeta_{7}^{}\right)\varepsilon
\nonumber\\&\quad
+\left( -\tfrac{34128}{5} \zeta_{3,5}^{} +\tfrac{1907604}{875} \zeta_{2}^{4} -12472 \zeta_{3}^{} \zeta_{5}^{}\right)\varepsilon^{2}
+\bigo{\varepsilon^{3}}

%% file: Expansions/M51-1111111111.tex
 -20 \zeta_{5}^{}\varepsilon^{-1}
 -188 \zeta_{3}^{2} -\tfrac{80}{7} \zeta_{2}^{3}
+\left( -\tfrac{1128}{5} \zeta_{2}^{2} \zeta_{3}^{} -751 \zeta_{7}^{}\right)\varepsilon
\nonumber\\&\quad
+\left( -294208 \zeta_{3}^{} \zeta_{5}^{} -\tfrac{814536}{5} \zeta_{3,5}^{} +\tfrac{46889448}{875} \zeta_{2}^{4}\right)\varepsilon^{2}
+\bigo{\varepsilon^{3}}

%% file: Expansions/M51-full.tex
-20 \zeta_5^{}\varepsilon^{-1}
-\tfrac{80}{7}\zeta_{2}^{3}
-\zeta_3^{2} \left( 68+6\,p_1 \right)
\nonumber\\&\quad
-\varepsilon \left\{
		\tfrac{1}{5} \zeta_2^{2}\zeta_3^{} \left( 408+36\,p_1 \right)
		+\zeta_7^{} \left( 170 - 7\,p_2 \right) 
\right\} 
+\bigo{\varepsilon^2}

%% file: Expansions/M51-p_1.tex
2\,\epe_{36810}+3\,\epe_{4579}

%% file: Expansions/M51-p_2.tex
8 \epe_{12}
-\tfrac{55}{4} \epe_{4579}
-\tfrac{5}{2} \epe_{36810}
-\tfrac{1}{8} p_1^2
+2 \left( \epe_{8}-\epe_{10} \right)  \left( \epe_{4510}-\epe_{789} \right)
+2 \left( \epe_{3}-\epe_{6} \right)  \left( \epe_{567}-\epe_{349} \right)
\nonumber\\&\quad
+2 (\epe_{12}\epe_{345678910}
		+\epe_{36}\epe_{810}
		-\epe_{47}\epe_{59} )
-4 \left( \epe_{4}^{2}
		+ \epe_{5}^{2}
		+ \epe_{7}^{2}
		+ \epe_{9}^{2} \right)